\documentclass[a4paper, final, twocolumn, accepted=2023-05-26]{quantumarticle}

\pdfoutput=1 %dont know why, needed by the quantum class
\usepackage{bbm} %bbm characters in math mode
\usepackage{amsmath} %enhancements to math mode
\usepackage{amssymb} %extended symbol collection
\usepackage{physics} %ket class among other things
\usepackage[bookmarks=true]{hyperref} 
\usepackage[english]{babel} 
\usepackage[normalem]{ulem} 
\usepackage[numbers,sort&compress]{natbib}

\usepackage{array}%Necesario para la tabla de fases
\usepackage{tikz}%needed for the appendix drawing
\usepackage{graphicx}
\usepackage[caption=false]{subfig}
\usepackage[]{xcolor} % allows access to named colors

\begin{document}
\title{Geometric phases along quantum trajectories}
\author{Ludmila Viotti}
\affiliation{Departamento de F\'isica Juan Jos\'e Giambiagi,
FCEyN UBA Ciudad Universitaria, Pabell\'on I, 1428 Buenos Aires, Argentina}
\affiliation{The Abdus Salam International Center for Theoretical Physics, Strada Costiera 11, 34151 Trieste, Italy}

\author{Ana Laura Gramajo}
\affiliation{The Abdus Salam International Center for Theoretical Physics, Strada Costiera 11, 34151 Trieste, Italy}

\author{Paula I. Villar}
\affiliation{Departamento de F\'\i sica {\it Juan Jos\'e
 Giambiagi}, FCEyN UBA and IFIBA CONICET-UBA, Facultad de Ciencias Exactas y Naturales,
 Ciudad Universitaria, Pabell\' on I, 1428 Buenos Aires, Argentina}

\author{Fernando C. Lombardo}
\affiliation{Departamento de F\'\i sica {\it Juan Jos\'e
 Giambiagi}, FCEyN UBA and IFIBA CONICET-UBA, Facultad de Ciencias Exactas y Naturales,
 Ciudad Universitaria, Pabell\' on I, 1428 Buenos Aires, Argentina}

\author{Rosario Fazio}
\affiliation{The Abdus Salam International Center for Theoretical Physics, Strada Costiera 11, 34151 Trieste, Italy}
\affiliation{Dipartimento di Fisica, Universit\`a di Napoli "Federico II'', Monte S. Angelo, I-80126 Napoli, Italy}

%\date{05/26/2023}                              

\begin{abstract}   
A monitored quantum system undergoing a cyclic evolution of the parameters governing its Hamiltonian accumulates a geometric phase that depends on the quantum trajectory followed by the system on its evolution. The phase value will be determined both by the unitary dynamics and by the interaction of the system with the environment. Consequently, the geometric phase will acquire a stochastic character due to the occurrence of random quantum jumps. 
Here we study the distribution function of geometric phases in monitored quantum systems and discuss when/if different quantities, proposed to measure geometric phases in open quantum systems, are representative of the distribution. We also consider a monitored echo protocol and discuss in which cases the distribution of the interference pattern extracted in the experiment is linked to the geometric phase. Furthermore, we unveil, for the single trajectory exhibiting no quantum jumps, a topological transition in the phase acquired after a cycle and show how this critical behavior can be observed in an echo protocol. For the same parameters, the density matrix does not show any singularity. We illustrate all our main results by considering a paradigmatic case, a spin-1/2 immersed in time-varying a magnetic field in the presence of an external environment. 
The major outcomes of our analysis are however quite general and do not depend, in their qualitative features, on the choice of the model studied.
\end{abstract}
%-------------------------------------------
\maketitle

\section{Introduction}\label{sec:intro}

As Berry first stated in his seminal work~\cite{Berry}, when a quantum system is prepared in an energy eigenstate and adiabatically driven in a cycle, it acquires, in addition to the dynamical phase, a phase that depends solely on the path traced in the ray space. Being independent of the specific dynamics giving rise to the path, this phase is of geometrical nature. Following Berry's breakthrough, consistent generalizations of the Geometric Phase (GP) have been found for unitary evolutions which are kept cyclic while they are not required to be adiabatic~\cite{aharonovanandan}, in the presence of degenerate subspaces~\cite{wilczeknonabelian}, and for the case in which both the adiabaticity and the cyclicity conditions are removed~\cite{samuelbhandari, mukunda93}. Further generalizations include the definitions of GPs for mixed states~\cite{uhlmann1986, uhlmann1989, uhlmann1991, Sjoqvistmixed, tongmixed} and the so-called off-diagonal GPs~\cite{maninioffdiagonal,sjoqvistoffdiagonal}, which apply in the case where the initial and final states are orthogonal. 

GPs are profoundly linked to the theory of fiber bundles and holonomies, bridging geometrical concepts like parallel transport over curved spaces with physics~\cite{simonmath,
nakaharamath, bohmmath}, and contributing in this way to the understanding of quantum mechanics at the foundational level. Since their discovery, GPs have also emerged in most diverse physical systems~\cite{book_phases, wilczek_book}, deepening the comprehension of numerous phenomena such as integer quantum Hall effect~\cite{thouless1982_app_hall}, topological insulators and superconductors~\cite{bernevig2013_app_supercond, asboth2016_app_supercond}, as well as playing a pivotal role in quantum information processing~\cite{Zanardi_1999,Jones_2000,nayak2008_app_qi}.

The quest for implementations of geometric quantum information processing has also spurred the search for geometric interferometry in several different setups. The first proposal of this kind was realized in NMR~\cite{Jones_2000}. Thereafter, Berry phases in superconducting qubits were both studied theoretically in~\cite{Falci_2000}  and observed experimentally for different regimes of couplings in circuit-QED arrangements~\cite{leek2007_cqed_observation, mottonen2008_cqed_observation, gasparinetti2016_cqed_observation, 
abdumalikov2013_app_qi_cqed,song2017_app_qi_cqed, xu2020_app_qi_cqed}.  In this direction, high-fidelity quantum gates were demonstrated with trapped ions~\cite{leibfried2003_app_qi}. 
The need to improve the performance of quantum information processing devices against the exposure to external environment has led to the suggestion of non-adiabatic geometric gates 
schemes~\cite{xiang2001_app_qi, zhu2002_app_qi, li2020_path,ding2021_path, sjoqvistshortcut, measuringshortcut}. 
In this context, it becomes of fundamental importance to understand how geometric interferometry is affected by the presence of an external environment. Consequently, GPs need to be generalized to deal with the systems subject to non-unitary quantum evolution. The effect of fluctuations in the classical control parameters of a quantum cyclic evolution may average out mitigating their effect on the accumulated Berry phase~\cite{De_Chiara_2003}. The presence of an external bath was found to give rise to new geometric contributions to decoherence~\cite{gefen2003_nonisolated, gefen2005_environment}, as experimentally detected in~\cite{ berger2013_noise_cqed, berger2015_noise_cqed}. 
Different definitions of GPs applicable in the non-unitary case have been put forward. Tong {\em et al.}~\cite{Tong_kinematic} introduced a purification-independent formula computed over the reduced density matrix while an average over different histories (trajectories) taking into account system-bath interaction was discussed in Carollo {\em et al.}~\cite{Carollo_original, Carollo_review} and further analyzed in~\cite{buri, Sjo_no, bassi2006_no}. Additional work along these lines can be found 
in~\cite{de_oldphases, ericsson_oldpahses, lombardo_villar06, lombardo_villar14}. 

There is, however, a different level of description of open quantum systems which may capture features that are washed out by simply looking at the properties of density matrices. 
This level is accessed, for example, when the state of the system is continuously monitored. In this case, the quantum system is described by a wave function whose smooth evolution is interrupted by random quantum jumps induced by the coupling with the environment~\cite{Molmer:93}. This sequence of smooth evolutions interrupted by jumps is named a quantum trajectory (see~\cite{manzano} for a recent review on the subject). There are also formal connections between this setting and that emerging in monitored quantum circuits~\cite{fisher2023random}, a tractable setting to explore universal collective phenomena as dynamics of quantum information and entanglement~\cite{kelly2022coherence, weinstein2022scrambling}.

{\em Goal of the present work is to describe the properties of accumulated GP along quantum trajectories.} In this approach we are inspired by the work of Gebarth {\em et al.}~\cite{gefenWeak} where the GPs induced by a sequence of weak measurements stirring the system along a path in a parameter space were analyzed. The randomness introduced by the occurrence of jumps in a given trajectory is reflected in the fact that the GPs inherit a stochastic nature. By random sampling over the trajectories, the entire distribution can be reconstructed. Since the Berry phase is not an observable, the average value does not correspond to the phase accumulated by the average state (this is, the density matrix). Previous works, with the notable exception of~\cite{gefenWeak}, either restrict the study of the dynamics of smoothly evolving pure states with no jumps or define average quantities. Understanding the fluctuations of GPs induced by random jumps is to a large extent unexplored. We would like to fill this gap by studying this distribution and whether it is related to the corresponding distribution in the interference fringes in a spin-echo experiment. Finally, with regard to the topological transition discussed in~\cite{gefenWeak}, further investigated theoretically in~\cite{snizhko2021weak, snizhko2021weak2} and experimentally observed in~\cite{wang2022observing, ferrer2022topological}, we will argue that despite the different dynamical settings it is a generic feature present in adiabatically driven monitored systems. We will show that depending on the coupling to the external environment, the monitored quantum system will show a topological transition in the phase accumulated in a cycle and we will argue that this transition is visible in echo dynamics.

The paper is organized as follows. In the next Section, we will define the dynamical setting we are interested in: A quantum system subject to a time-periodic Hamiltonian and coupled to an external bath. With the intention to highlight the essence of our results, we will consider the paradigmatic case of a two-level system that evolves in presence of an externally varied magnetic field. The associated density matrix is governed by the Lindblad equation. In order to follow the dynamics of the system along its quantum trajectories, we introduce a specific unravelling of the Lindblad equation which relays on microscopic considerations, these aspects are introduced in Section~\ref{sec:th_qtraj}.  In Section~\ref{themodel} the model and its coupling to the environment are introduced. In Section~\ref{sec:th_gp} we define the GP that will be the founding block of all our analysis. For an isolated system and sufficiently slow driving, this reduces to the Berry phase~\cite{Berry}. The presence of the environment induces both a smooth drift and random jumps in the dynamics, so the evolution of the state is generically neither adiabatic nor cyclic. To keep the presentation self-consistent, we further include in this same Section other definitions of GPs present in the literature. These will be employed for comparison in the posterior Section~\ref{sec:unrav_gp}, where we discuss the distribution of the GPs accumulated along quantum trajectories and analyze reference GP values in order to account for differences with other definitions of GPs proposed in the context of open quantum systems. Due to the intrinsic randomness of the quantum trajectory, a monitored echo experiment might be altered. In Section~\ref{sec:unrav} we discuss the probability distributions of the interference fringes and detail whether/when they relate to the corresponding distribution of the GPs. Our analysis of GPs in monitored systems is completed in Section~\ref{sec:topological} where we will show that the topological transition discovered in~\cite{gefenWeak} for a specific setting is actually a generic feature in periodically driven open quantum systems. Indeed, for the sequence of states known as no-jump trajectory, which can be thought of as the smooth evolution generated by a non-hermitian Hamiltonian, we find the GP displays a complex pattern in the parameters space exhibiting singular points. 
These singularities can be tracked down to correspond to points of vanishing probability for such a trajectory, and to reveal the border between distinct topological sectors. The transition observed in the evolution when varying the parameters is topological in the sense that it is related to a discontinuous jump of an integer-valued topological invariant. Section~\ref{sec:topological} will be entirely devoted to the study of this transition and ways to detect it through an echo protocol. A summary of our results and concluding considerations are presented in Section~\ref{sec:concl}. 
The appendices give some additional ingredients used to compute the GP in the numerical simulations, Appendix~\ref{ap:gp_derivation}, a detailed analysis of the already mentioned interference fringes distribution, Appendix~\ref{ap:echo_distribution},  a brief discussion on how the distribution of GPs may depend on the unravelling of the Lindblad equation (leading to the same averaged evolution), Appendix~\ref{ap:diff_unrav}, and analytical treatment of the no-jump trajectory, Appendix~\ref{ap:analytic}.
  
\section{From Lindblad dynamics to quantum trajectories}\label{sec:th_qtraj}

{\em Lindblad equation - } In order to make a connection with existing literature, it is convenient to set the stage and start from the case in which the state of an open quantum system is described by a density matrix $\rho (t)$. In this case, under proper conditions, the dynamics is governed by the Lindblad equation~ \cite{lindblad1976,rivas2012_book} ($\hbar =1$)

\begin{equation}
   \dot{\rho} = -i\left[H, \rho \right] + \sum_{\alpha} [L_{\alpha}\rho L_{\alpha}^\dagger - \frac{1}{2} \{L_{\alpha}^\dagger L_{\alpha}, \rho\, \} ] \;.
   \label{eq:Lindblad}
\end{equation}
The first term in the r.h.s. of the Lindblad equation accounts for the unitary evolution, while the second originates in the coupling to the environment. The strength and the nature of this coupling are encoded in the Lindblad operators $L_{\alpha}$. We will consider a Hamiltonian $H$ that depends periodically on time $H(t+2\pi/\Omega) = H(t)$ with 
$T=2\pi/\Omega$ the period of a cycle in suitable parameter space. The Lindblad operators, if time-dependent, should also be time-periodic $L_{\alpha} (t+2\pi/\Omega)= L_{\alpha} (t)$.

It is useful to already at this point briefly comment on the adiabatic limit for slow dynamics as this issue will be central in the analysis conducted along the paper. If the evolution is unitary, for a sufficiently large period $T$, a system prepared in an eigenstate will remain in the corresponding instantaneous eigenstate up to small corrections due to Landau-Zener transitions between energy levels. In other words, the occupancy of any given eigenstate will not change in time. The situation strongly differs in presence of an environment. In this case, a proper adiabatic limit is not well defined, since the slow driving limit where adiabatic dynamics sets in, is also the regime in which the consequences of the external baths are the most severe and the system reaches a (possibly periodic) steady state. The adiabatic limit itself should be reconsidered~\textcolor{red}{\cite{Sarandy_2005, thunstrom2005adiabatic, yi2007adiabatic, oreshkov2010adiabatic, venuti2016adiabaticity}} in  an open system, as the existence of a continuum of energy levels makes the energy splittings of the system a bad reference scale for defining the regimes. Effects due to non-adiabaticity and corrections due to the presence of the environment seem thus to be inextricably linked. 

{\em Monitored dynamics and quantum trajectories - }
The dynamics of the systems radically change when it is possible to continuously monitor their state. In this case, the state of the system remains pure and consists of intervals of 
smooth evolution interrupted at random times by abrupt changes called quantum jumps. A sequence of smoothly-evolving intervals together with a set of random events is denominated 
a quantum trajectory. The literature on the subject is vast and we refer to the following papers and books for a general overview~\cite{Molmer:93, carmichael1993_open, wiseman2009quantum, manzano} and applications e.g. to many-body systems~\cite{daley2014quantum, passarelli2019improving}. 

Evolution is described in this framework as follows. If at time $t$ the state of the system is $|\psi(t)\rangle$, at a later $t+\delta t$ time it will be

\begin{equation}
|\psi(t+\delta t)\rangle =  \left\{
\begin{tabular}{ ccc } 
 $ \frac{K_o|\psi(t)\rangle}{\sqrt{p_{o}(t)}} $ & with probability  &$p_o(t)$ \\
          &         &         \\ 
 $ \frac{K_{\alpha} |\psi(t)\rangle}{\sqrt{p_{\alpha}(t)}} $ & with probability   &$p_{\alpha}(t)$\\
 \end{tabular}
 \right.
 \label{eq:monitored-evol}
\end{equation}
where $o, \alpha = 1, ..$ label the different operators $K_\alpha$ inducing dynamical steps

\begin{equation}
    K_{o} = 1- i\,\delta t \left[H-\frac{i}{2}\sum_{\alpha} L^\dagger_{\alpha}L_{\alpha}\right]  \;\;\;\;\;\;\;\;    K_{\alpha} = \sqrt{\delta t}L_{\alpha}
    \label{eq:jumpoperators}
\end{equation}
and $p_{o/\alpha}(t) = \bra{\psi(t)}K_{o/\alpha}^\dagger K_{o/\alpha} \ket{\psi(t)}$. Each choice in the r.h.s. of Eq.(\ref{eq:monitored-evol}) represents evolution steps of different characters. 
The second line corresponds to the occurrence of a jump $K_{\alpha}$ at time $t$, while the first is a smooth evolution (no jump), albeit altered from unitarity by the fact that acquiring the information that no jumps occurred modifies the evolution of the system. The no-jump operator $K_{o}$ can also be thought of as generated by an effective drift Hamiltonian $H_o$ to which it relates in the usual way $K_{o} = 1 - i \, \delta t\, H_{o}$. The full evolution in a time interval $[0, t]$ is therefore characterized by a sequence of $N_J$ jumps of types $\alpha_i$ occurring at times $t_i$. We will denote the string of these events

\begin{equation}
	{\cal R} (t, N_{J}) = \{(\alpha_1, t_1), \dots, (\alpha_i, t_i),\dots(\alpha_{N_J}, t_{N_{J}})\},
	\label{eq:R}
\end{equation}
with $0 \ge t_i \ge t \;\; \forall  i$, the quantum trajectory.
As mentioned above, this framework naturally emerges when the system is continuously and indirectly monitored, so that each trajectory can be viewed as the result of continuous measurements of the environment on a given basis. From this perspective, continuous monitoring may lead to decoherence mitigation by the environment~\cite{siddiqi2_observing},
also post-selection and error correction schemes~\cite{ahn2002_error, siddiqi1_observation} have been proposed. 

The properties of the Kraus operators $K_{o/\alpha}$ guarantee that the probabilities to get a given outcome sum up to one, and the time step $\delta\,t$ should be taken small enough for the first order approximation to be valid, which requires $\sum_{\alpha } p_{\alpha} \ll 1$.  Averaging over every possible jump sequence one gets back the Lindblad equation~\cite{Molmer:93} in Eq.(\ref{eq:Lindblad}), the converse implication is not valid, an infinite number of different unravellings give rise to the same Lindblad evolution~\cite{manzano}. 
We will address this question in Appendix~\ref{ap:diff_unrav}.

\section {The model}
\label{themodel} 
Since we are interested in studying the impact of an external environment on the GPs, we will consider a unitary evolution over which the accumulated GP, in the adiabatic limit, is the Berry phase. To be concrete, we shall consider a spin-1/2 particle in presence of a time-dependent magnetic field $\mathbf{B}(t) =\omega\, \hat{\text{\bf{n}}}_\mathbf{B}(t)$, whose direction is given by $\hat{\text{\bf{n}}}_\mathbf{B }=(\sin{(\theta)}\cos(\Omega\,t),\sin{(\theta)}\sin(\Omega\,t), \cos{\theta})$ with fixed polar angle $\theta$ and time-varying azimuthal angle $\Omega\,t$. 
Such unitary evolution is generated by the Hamiltonian

\begin{equation}
    H(t) = \frac{1}{2}\,\mathbf{B}(t)\cdot \boldsymbol{\sigma},
    \label{eq:Hamiltonian}
\end{equation}
with $\boldsymbol{\sigma} = (\sigma_x, \sigma_y, \sigma_z)$; and $|0\rangle$ and $ |1\rangle $, the eigenstates of $\sigma_z$. The instantaneous eigenstates of $H(t)$ are denoted $\ket{\psi_-(t)}$ and $\ket{\psi_+(t)}$ .
 
\begin{figure}[ht!]
    \centering
    \includegraphics[width = .6\columnwidth]{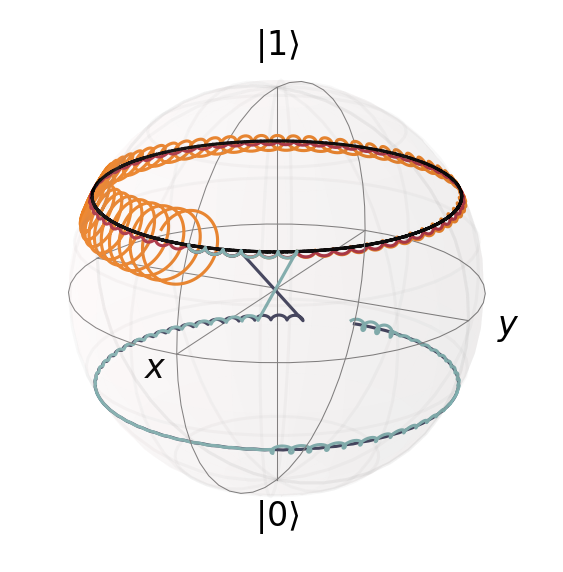}
    \caption{Trajectories described by the state of the system on the Bloch sphere under different conditions. The black line corresponds to unitary evolution in the adiabatic limit. 
    The purple line depicting a curly ring corresponds to general unitary dynamics in which non-adiabatic corrections start to be visible. In the presence of an environment, the quantum state can suffer from jumps or can be smoothly driven along the whole evolution. For a system prepared in the exited eigenstate, the orange trajectory corresponds to a fully smooth drift. Differently, the blue path shows a jump that projects the state into the instantaneous ground eigenstate and is afterward smoothly driven. Finally, the light blue path shows a case with several jumps, where the non-adiabatic corrections appear in between the jumps.}
    \label{fig:bloch}
\end{figure}

If the system could be kept perfectly isolated while the direction of $\mathbf{B}(t)$ is adiabatically changed in a cycle parameterized by $t \in [0, T]$, with $T = 2\pi/\Omega$ (as shown in Fig.\ref{fig:bloch}), it would acquire an adiabatic (Berry) phase $\phi^{\pm}_{\mbox{{\small a}}} = -\pi(1\mp\cos\theta)$, where the $\mp$ sign depends on the energy eigenstate in which the system was initially prepared. 

{\em Lindblad operators - }
For a system that evolves according to $H(t)$ given by Eq. (\ref{eq:Hamiltonian}) coupled to an environment of harmonic oscillators a consistent time-dependent Lindblad equation 
of the form in Eq. (\ref{eq:Lindblad}) can be derived from microscopic considerations as long as the evolution remains sufficiently slow~\cite{albash2012, albash2015}, with Lindblad operators given by

\begin{eqnarray}
	L_{-}(t)  &= &\sqrt{\gamma_-}\bra{\psi_-(t)}\sigma_x\ket{\psi_+(t)}\ket{\psi_-(t)}\bra{\psi_+(t)} \nonumber\\  
	L_{+}(t) &= &\sqrt{\gamma_+}\bra{\psi_+(t)}\sigma_x\ket{\psi_-(t)}\ket{\psi_+(t)}\bra{\psi_-(t)}\label{eq:operators}\\
	L_{d}(t) &=  &\sqrt{\gamma_d}\sum_{i=\pm}\bra{\psi_i(t)}\sigma_x\ket{\psi_i(t)}\ket{\psi_i(t)}\bra{\psi_i(t)}  \nonumber
\end{eqnarray}
and corresponding to decay, spontaneous excitation, and dephasing respectively. The coupling strengths considered in this work are, in terms of the dissipation ratio $\Gamma$, $\gamma_-=\Gamma\;;\;\gamma_d=0.32\, \Gamma$, while we consider $\gamma_+$ to be negligible (all the results that we will show are rather generic and do not qualitatively depend 
on the chosen values). The jumps defined by Eq. (\ref{eq:jumpoperators}) from the Lindblad operators above lead, after averaging, to a consistent Lindblad equation for slow dynamical evolutions~\cite{jumpsAdiabatic}. The operators introduced in Eq. (\ref{eq:operators}) induce transitions and dephasing between the instantaneous eigenstates of the Hamiltonian defined in 
Eq.(\ref{eq:Hamiltonian}). In order to keep the analysis as general as possible, we will include a further term in the Lindbladian which requires considering a fourth operator

\begin{equation}
    L_{z} =  \sqrt{\gamma_z} \sigma_z
    \label{eq:operator_z}
\end{equation}
along a fixed direction in the Bloch sphere. The particular choice of $\sigma_z$ operator as the additional Lindblad operator is motivated by the need of introducing transitions that do not simply involve the instantaneous eigenstates. Any other Lindblad operator that differed from those in Eq.(\ref{eq:operators}) would lead to similar qualitative conclusions. 

While the unitary evolution of the closed system will follow the curly path indicated in purple in Fig.\ref{fig:bloch}, the actual dynamics will follow, with some probability, the path indicated in blue (see Fig.\ref{fig:bloch} for illustrative purposes), i.e. it will be discontinuous and not necessary closed after a cycle of the driving, even in the slow-driving limit. Moreover, the slower the driving, the more jumps will occur (see light blue curve in Fig.\ref{fig:bloch}). The task of the next Sections is to characterize GPs under these conditions.

{\em Smooth evolution with no jumps - } A particularly interesting quantum trajectory is that which is smooth along the whole evolution. Before addressing the characterization of GPs in indirectly 
monitored systems, we provide insight into the evolution giving rise to it. When the records of the measurements performed on the environment reveal zero jumps, the dynamics describe a continuous smooth path and is generated by an effective drift Hamiltonian which depends both on the Hamiltonian of the system and the Lindblad operators as described by Eq. (\ref{eq:jumpoperators}).
Within the model considered in our work, the effective drift Hamiltonian $H_o$ governing the no-jump dynamics [$K_o = 1 - \delta t H_o$ in Eq.(\ref{eq:jumpoperators})] is given by

\begin{equation}
H_o(t) = \left(1 - i\frac{\Gamma}{2\omega} f(t)\right)\,H(t)
\label{eq:nojump_hamiltonian}
\end{equation}
with $f(t)=\cos^2(\theta) + \sin^2(\theta)\sin^2(\Omega t)$. We highlight the fact that, due to the unitarity of $\sigma_z$ matrix, the no-jump evolution is completely independent of the fourth 
Lindblad operator $L_z$ included ad-hoc and, consequently, from the parameter $\gamma_z$.
An illustrative example of the trajectory generated by the above evolution, referred to as the no-jump trajectory in what comes, is the orange path in Fig. \ref{fig:bloch}. In Appendix \ref{ap:analytic} 
we provide the analytic solution for the dynamics associated with the non-Hermitian Hamiltonian $H_o(t) $ of Eq.(\ref{eq:nojump_hamiltonian}).
While this trajectory is unique, the number of possible (even though unevenly probable) trajectories in which $N_J > 0$ jumps occur increases with the number $N_J$ of jumps, diverging
as $\delta t$ goes to zero. Its uniqueness will make the no-jump trajectory especially suitable for the analysis of some features of the GPs,  we come back to this question in Section~\ref{sec:topological}. 

\section{Geometric phases in open systems}\label{sec:th_gp}

As mentioned in Sec. \ref{sec:intro}, the accumulation of a GP during the dynamics of a quantum system is not necessarily restricted to an adiabatic evolution. For a generic quantum trajectory, consisting of a sequence of smoothly-evolving intervals together with a set of random quantum jumps $\mathcal{R}$, a proper phase that deals with both aspects of evolution can be defined. 

Considering the evolution in a time interval $[0,T]$, parameterized with $t$, the GP associated to a trajectory in which $N_{J}$ jumps are registered at times $t_i$, can be written as 

\begin{eqnarray}\nonumber
    \phi [{\cal R}] = & & \arg\bra{\psi(0)}\ket{\psi(T)}   \nonumber \\
	   	&-& \text{Im}\sum_{i=0}^{N_{J}}\int_{t_i}^{t_{i+1}}\hspace{-0.1cm} dt\;\frac{\bra{\psi(t)}\dot{\psi}(t)\rangle}{\bra{\psi(t)}\ket{\psi(t)}} \nonumber \\
		&-& \sum_{(t_i, \alpha_i) \in \mathcal{R}} \arg\bra{\psi(t_{i})}K_{\alpha_i}\ket{\psi(t_{i})} ,
    \label{eq:GP_traj}
\end{eqnarray}
where ${\cal R} = {\cal R}(T, N_J)$ for brevity, with $t_0 = 0$ and the convention that  $t_{N_{J} +1} \equiv T$ in the sum of integrals. 
The definition of GP as given in Eq.(\ref{eq:GP_traj}) will be at the basis of our analysis and we refer to Appendix \ref{ap:gp_derivation} for a derivation of this expression. 
As it is evident from the dependence on the times and nature of the jumps, the phase $\phi [{\cal R}(T, N_J)] $ will be a stochastic variable, dependent on the trajectory ${\cal R}(T, N_J)$.
The first term in Eq. (\ref{eq:GP_traj}) is the total relative phase between the initial and final states. The remaining terms are of two different kinds, reflecting the properties of the dynamics itself. The second term features the dynamical phases accumulated along the intervals of smooth evolution that take place before, between, and after jumps, and which should be subtracted in order to access the purely geometrical object $\phi_{\cal R}$. The occurrence at time $t_i$ of a jump generated by the operator $K_{\alpha_i}$ introduces a contribution 
$\arg\bra{\psi(t_{i})}K_{\alpha_i}\ket{\psi(t_{i})}$ given by the Pancharatnam phase difference $\phi_P^{1\rightarrow 2} =\arg\bra{\psi_1}\ket{\psi_2}$ between the state before and after the jump. Such terms, arising when the path traced by the state shows discontinuities~\cite{mukunda93, samuelbhandari}, can also be thought of as the GP accumulated along the geodesic arc in the Hilbert space that fills the discontinuity by joining both states. The expression in Eq. (\ref{eq:GP_traj}) is independent of the $U(1)$ gauge choice. It neither requires 
the trajectory to trace a close path in the state space nor relies on adiabaticity condition. Moreover, it does not even demand unitarity as it is well defined also if the states $\ket{\psi(t_i)}$ or $\ket{\psi(t')}$ are not normalized (the norm should, however, be non-vanishing).

Suitable to be applied to the trajectories that emerge in master equation unraveling, Eq. (\ref{eq:GP_traj}) has been employed in limiting forms for addressing the definition of GPs fitting non-unitary evolution. A first explored route was to focus on the no-jump trajectory~\cite{Carollo_original, Carollo_review}. This approach, which disregards the possibility of quantum 
jumps by restricting to the smooth evolution, preserves the well-known definitions of GPs applicable to pure states and includes environmental effects through the non-hermiticity of $H_o$. If no jumps are registered along the entire evolution, this is, if $\mathcal{R}(T, 0) = \emptyset$, the GP $\phi_{0} \equiv \phi [{\cal R}(T, 0)= \emptyset] $ reads

\begin{equation}
    \phi_{0}  = \arg\bra{\psi(0)}\ket{\psi(T)}  -\text{Im}\int_{0}^{T}\;dt\;\frac{\bra{\psi(t)}\dot{\psi}(t)\rangle}{\bra{\psi(t)}\ket{\psi(t)}}
    \label{eq:GP_nj}
\end{equation}
which trivially reduces to the expression for the GP accumulated in the most general unitary evolution~\cite{mukunda93} when this is indeed the case, and therefore the states are instantaneously normalized, rendering the denominator $\bra{\psi(t)}\ket{\psi(t)}\equiv 1\; \forall \, t$. Eq.(\ref{eq:GP_nj}) also reduces to Aharonov-Anandan and Berry phases as the conditions required by each definition are fulfilled, namely, for cyclic and unitary while not necessarily adiabatic evolution and for both cyclic and adiabatic evolution. Note that phase $\phi_{0}$ is ill-defined if some internal product on its argument vanishes, this observation will become of relevance when discussing the topological transition in Section~\ref{sec:topological}. 

Several other works consider the full Lindblad equation unraveling, suggesting to define the GP of the ensemble-averaged state $\rho(t)$ as an average over the ensemble of phases $\{\phi_{\mathcal{R}}\} =\phi_{\{\mathcal{R}\}}$ obtained by applying Eq.(\ref{eq:GP_traj}) to each trajectory \cite{Carollo_original, Carollo_review, buri}. It has been extensively discussed 
whether this is a proper definition of a GP for the density matrix representing the state of the system as it does not allow for a one-to-one relation between the set of density matrices and the obtained GP values \cite{Sjo_no, bassi2006_no, sjoqvist2010_hidden}.

Finally, a different approach introduces a generalized GP defined directly from the reduced density matrix~\cite{Tong_kinematic}. 
The expression reads
\begin{align}
\phi_{\rho} = \arg\left(\sum_j \sqrt{ \lambda_m (0) \lambda_m (t)}
\bra{\xi_m(0)}\ket{\xi_m(t)} \right. \nonumber \\
\times  \left.  \exp{-\int_0^{t} dt' \bra{\xi_m(t')}
\Dot{\xi}_m(t')\rangle} \right)\;\; \label{eq:phiNonU}
\end{align} 
where $\lambda_k(t)$ and $|\xi_k\rangle$ are the instantaneous eigenvalues and eigenstates of the density matrix $\rho(t)$ which describes the state of the system. Even though defined for non-degenerate but otherwise general mixed states, when computed over pure states under unitary evolution, reduces to the unitary expression of the GP. 

All the above-mentioned proposals of GPs applicable when dynamics are non-unitary either restrict to modified evolutions on which pure-state GP definitions would be applicable or seek a consistently defined GP for the reduced density matrix $\rho(t)$, which accounts for an averaged description.
Stochastic processes, however, arising from master equation unraveling, acquire independent physical relevance in continuous monitoring schemes. 
As anticipated in the introduction, the randomness introduced by the occurrence of jumps in a given trajectory reflects in the GPs acquiring a stochastic nature itself. This approach, therefore, requires a study of the environmentally-induced effects in GPs from a statistical perspective.
The probability associated with some GP value will be related to that of individual trajectories as
\begin{equation}
    P[\phi] = \sum_{\mathcal{R}/\phi[{R}]=\phi} P[\mathcal{R}].
    \label{eq:phase_probability}
\end{equation}
The average phase corresponds only to the first moment of the distribution
\begin{equation*}
    \Bar{\phi} = \arg\left(\sum \,e^{i\phi} P[\phi]\right)
\end{equation*}
and in some cases may be not sufficient in characterizing the dynamics.

For easy later reference, we provide a table summarizing the GP definitions reviewed along this section

\begin{center}
\vspace{0.5cm}
    \begin{tabular}{|>{\centering\arraybackslash}p{.1\linewidth}|>{\arraybackslash}p{.62\linewidth}|>{\centering\arraybackslash}p{.15\linewidth}|}
    \hline
    GP & {\hspace{2cm} Description}  & \\
    \hline
    $\phi_{\mathrm{a}}$& Adiabatic Berry phase &\\
    $\phi[{\cal R}]$ & GP associated to the quantum trajectory $\mathcal{R}(T, N_J)$&  Eq.(\ref{eq:GP_traj})\\
    $\phi_{0}$ & GP associated to the no-jump trajectory & Eq.(\ref{eq:GP_nj})\\
    $\phi_{u}$ & GP accumulated on general unitary evolution ( from Eq. (\ref{eq:GP_nj}) with $\bra{\psi(t)}\ket{\psi(t)}=1$) & \\
    $\Bar{\phi}$ & Average over the probability distribution $P[\phi]$&\\
    $\phi_{\rho}$ & Mixed state geometric phase ~\cite{Tong_kinematic} & Eq.(\ref{eq:phiNonU})\\  
    \hline
\end{tabular}
\vspace{0.5cm}
\end{center}

The next Section will be devoted to the properties of $P[\phi]$ and how representative the different GPs applicable to trajectories are, see Eqs. (\ref{eq:GP_traj} - \ref{eq:GP_nj}). As we will be showing in the following, in most cases the entire probability distribution, i.e. all higher order cumulant, is necessary to understand the accumulation of GPs in a continuously 
monitored system. We will also discuss under which circumstances and what features of $P[\phi]$ can be extracted by geometric interferometry through a spin-echo protocol.

\section{Results}
\subsection{Geometric phase distribution \texorpdfstring{$P[\phi]$}{Pp}} \label{sec:unrav_gp}
\begin{figure}[!ht]
    \centering 
    \includegraphics[width =.97\columnwidth, trim={0 0 .5cm 0}]{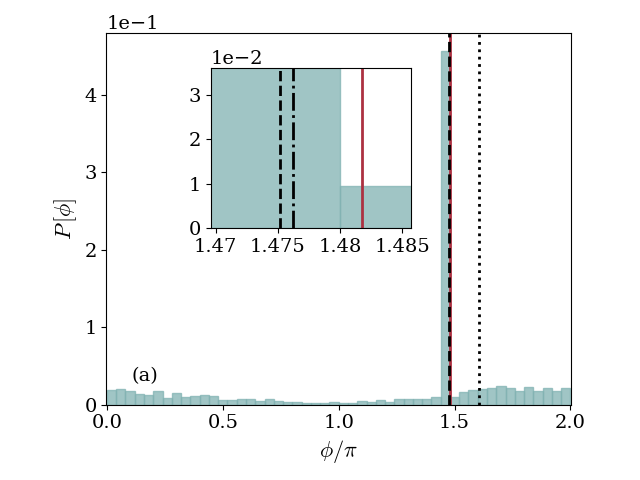}
    \includegraphics[width =\columnwidth]{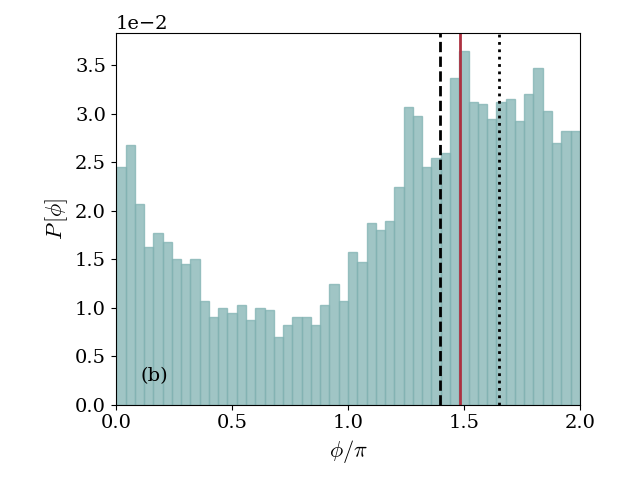}
    \caption{Probability distribution $P[\phi]$ of GPs for a magnetic field oriented with $\theta = 0.34\pi$ and driven in a loop at frequencies (a) $\Omega = 5\times 10^{-3}\omega$ 
    and (b) $\Omega = 5\times 10^{-4}\omega$. The environment is characterized by the dissipation rate $\Gamma = 10^{-3}\omega$ and a $\gamma_z = 0$. 
    In both panels, the solid red line depicts the adiabatic (Berry) phase $\phi^+_\mathrm{a}$, and the black dashed and dot-dashed lines signalize the GPs $\phi_0$ and $\phi_u$ associated with no-jump and general unitary evolution. The black dotted line indicates the first moment of the distribution $\Bar{\phi}$. 
    The inset in panel (a) is a zoom in which the difference between these reference GP values is visible.}
    \label{fig:hist_phi}
\end{figure}

We investigate in this section the distribution of the ensemble $\{\phi_{\mathcal{R}}\}=\phi_{\{\mathcal{R}\}}$ of GPs obtained by applying Eq. (\ref{eq:GP_traj}) to each individual 
realization (trajectory) of the evolution, characterized by some set $\mathcal{R}(T, N_J)$.
In Fig. \ref{fig:hist_phi} we show two representative cases in which the corresponding dynamics of a hypothetical {\em unitary} evolution would either be faster (with small but non-zero non-adiabatic corrections) or slow enough to be considered in the adiabatic regime while the environment remains the same, characterized by the dissipation rate $\Gamma = 10^{-3}\omega$, which leads to $\gamma_- = \Gamma$, $\gamma_d = 0.32\, \Gamma$, and negligible $\gamma_+$.

We first attend the case with $\gamma_z = 0$, in which the environment induces jumps involving instantaneous eigenstates only.  The two situations, corresponding to the two sets of parameters indicated before, are shown in Fig. \ref{fig:hist_phi}, in panels (a) and (b) respectively. In both panels, we also plot for reference the adiabatic (Berry) result, the no-jump and unitary GPs, and the average of the distribution. Being the Berry phase independent of $\Omega$, it is exactly the same for both cases, this is, $\phi_\mathrm{a} \sim 1.482 \pi$. For the parameters chosen, the value $\phi_{0}$ computed from Eq.(\ref{eq:GP_nj}) over the trajectory with no jumps, shows small deviations from $\phi_\mathrm{a}$. While the values of these characteristic GPs are similar, the 
entire distribution of the monitored system is drastically different on each panel. In the first case of faster driving, the period $T$ is such that a considerable amount of times the evolution is completed registering no jumps, with the mean number of jumps over the ensemble $\Bar{N}_J=0.63$. The narrow peak in the Figure shows these cases of entire smooth evolution. In addition, 
there is a small background revealing the accumulated GP along those trajectories where jumps occurred. 
The composition of the ensemble is reflected in the histogram by the presence of a large contribution, corresponding to $\sim 50\%$ of the realizations, due to the no-jump GP-value and the remaining $50\%$ of the counts distributed in a broad way over the possible GP values. This broad background distribution can be easily interpreted as the randomness inherited by the GP due to the (random) time at which the jump occurred. A single term $\bra{\psi(t_i)}K_{-_i}\ket{\psi(t_i)}$ in Eq. (\ref{eq:GP_traj}), denoting a contribution to the GP from a jump at time $t_i$, successfully accounts for the background when considering all possible jump-times. 
The peak in the distribution agrees well with both the adiabatic and the no-jump values. The average phase, on the other side, is a bit off due to the small and poorly structured background, broadly distributed over $2\pi$. This clearly demonstrates that even a single jump occurring at a random time leads to very large fluctuations in the accumulated GP.
In the case with slower driving shown in panel (b) the mean number of jumps over the set of trajectories is $\Bar{N}_J=1.77$. This means that the state of the system is much more likely to undergo an abrupt change, or even more than one, in each realization of the cycle. As expected, the distribution of GPs becomes much wider, and a sharp peak around $\phi_0$ is not visible anymore. Higher-order cumulants become necessary to understand the dynamics. The three lines, corresponding to the adiabatic, no-jump, and average GPs do not provide thorough information on the dynamics of the monitored system.

\begin{figure}[h]
    \includegraphics[width = \linewidth]{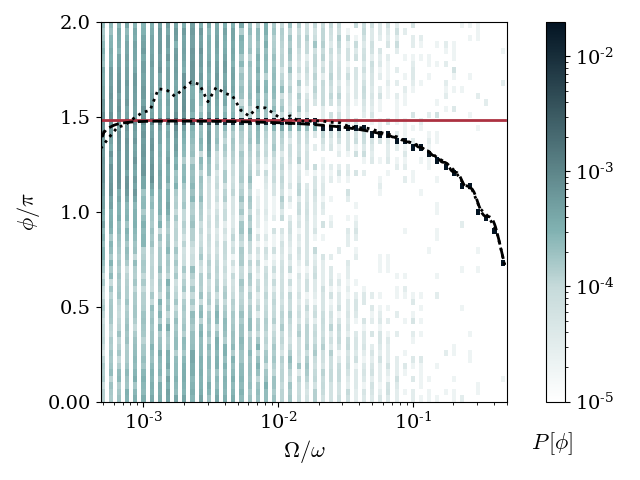}
    \caption{Probability distribution $P[\phi]$ of GPs as a function of the ratio $\Omega/\omega$. The field is oriented with $\theta = 0.34\pi$ and the environment is characterized by the 
    dissipation rate $\Gamma = 10^{-3}\omega$ and a $\gamma_z = 0$ amplitude for the fourth Lindblad operator. The GP values are displayed on the y-axis, while the intensity of the count color indicates their probability. The solid red line depicts the adiabatic (Berry) phase $\phi^+_\mathrm{a}$, the black dashed line indicates the GP $\phi_0$ accumulated along smooth trajectories with no jumps, and the black dotted line shows the first moment of the distribution $\Bar{\phi}$.}
    \label{fig:phi_vs_Omega}
\end{figure}

\begin{figure}[!ht]
    \centering 
    \includegraphics[width = .95\columnwidth]{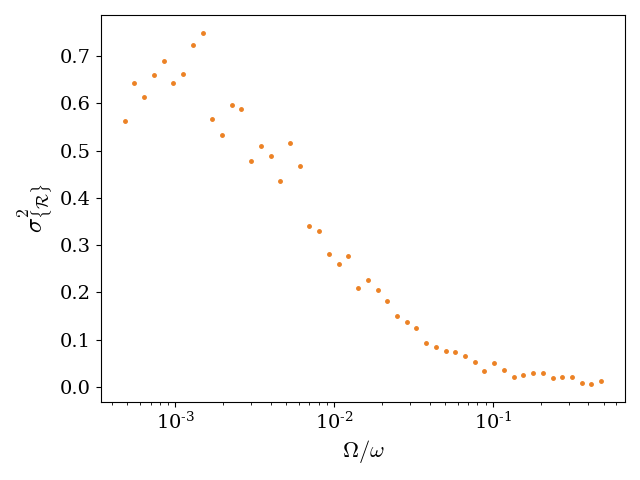}
    \caption{Variance $\sigma^2_{\{\phi_{\mathcal{R}}\}}$ of the GPs' distribution as a function of the ratio$\Omega/\omega$. The field is oriented with $\theta = 0.34\pi$ and the environment is characterized by the dissipation rate $\Gamma = 10^{-3}\omega$ and a $\gamma_z = 0$ (same as in Fig.\ref{fig:phi_vs_Omega}).}
    \label{fig:variance}
\end{figure}

The rate $\Omega/\omega$ at which the magnetic field is rotated has thus a direct impact on the distribution of GPs. For larger rates, the system is exposed to the environment for a shorter period of time, but deviations from the adiabatic regime become non-negligible. On the other hand, lowering the driving frequency might result in the system being exposed to environmental effects for too long, implying strong corrections to $\phi_{\mathcal{R}}$ from $\phi^+_\mathrm{a}$. Fig. \ref{fig:phi_vs_Omega} shows the distribution of GP-values obtained along a range of 
different $\Omega/\omega$ rates which include the cases presented in Fig.\ref{fig:hist_phi}.
For high enough frequency, the distribution shows a sharp peak around the no-jump value of the GP and almost no background counts. On the other hand, this no-jump value deviates considerably from the Berry phase. The broad background visible in panel (a) of Fig. \ref{fig:hist_phi} develops as the frequency rate is lowered and the relative period grows. Further on, the background turns into a second peak, while the one in the no-jump value decreases. For the smaller rate 
values, the distribution shows the behavior depicted by panel (b) of Fig. \ref{fig:hist_phi}, this is, a broad single-peaked distribution. This regime shows non-negligible environmental effects also over the GP associated with the no-jump evolution, which deviates from the adiabatic result even though the driving is performed slowly.
We refer to Appendix \ref{ap:analytic} for an analytical expression of the dependence of this deviation on the different parameters involved. The broadening exhibited by the distribution as the frequency rate decreases is reflected in the increment of the distribution variance. We note that, for variables described on a circle as the GP, this measure is constrained to the $[0,1]$ range. The described behaviour is shown in Fig. \ref{fig:variance}.

\begin{figure}[!ht]
    \centering 
    \includegraphics[width = .95\columnwidth]{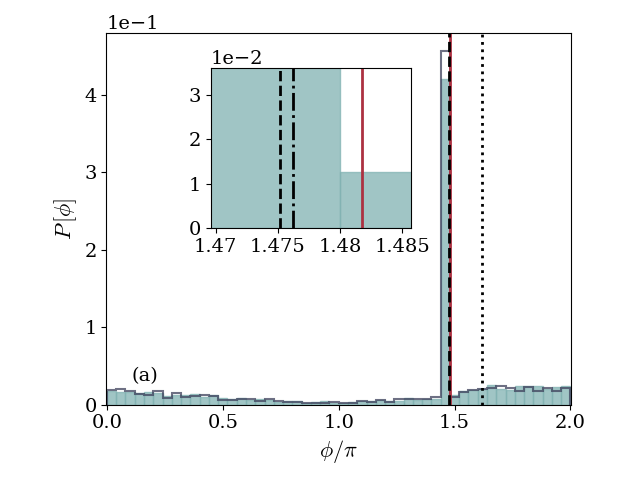}
    \includegraphics[width =\columnwidth]{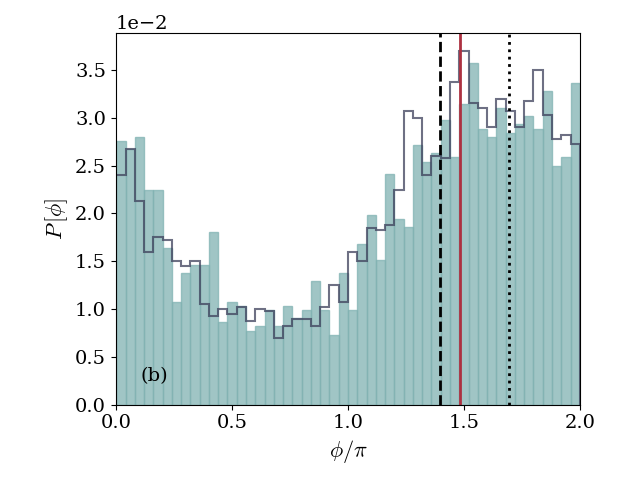}
    \caption{Probability distribution $P[\phi]$ of the GPs for a magnetic field oriented with $\theta = 0.34\pi$ and driven in a loop at frequencies (a) $\Omega = 5\times 10^{-3}\omega$ and (b) $\Omega = 5\times 10^{-4}\omega$. The environment is characterized by the dissipation rate $\Gamma = 10^{-3}\omega$ and  $\gamma_z = 0.1\,\Gamma$. In both panels, a blue solid contour indicates (for comparison) the $\gamma_z=0$ distributions. The solid red line depicts the adiabatic (Berry) phase $\phi^+_\mathrm{a}$, the black dashed and dot-dashed lines signalize the GPs $\phi_0$ and $\phi_u$ associated with no-jump and general unitary evolution. The black dotted line shows the first moment of the distribution $\Bar{\phi}$. The inset in panel (a) zooms in to see the differences in the positions of the lines}
    \label{fig:hist_phi_gz04}
\end{figure}

We conclude this section by analyzing the distribution of GP values when $\gamma_z \ne 0$. As already discussed, a non-zero value of $\gamma_z$ induces jumps to states that are not instantaneous eigenstates of the Hamiltonian and thus allows to consider of a wider class of cases. The resulting phenomenology depends only quantitatively on the choice of the Lindblad operator $L_z$.
Specifically, we take $\gamma_z = 0.1\,\Gamma$ and consider, as we did before, two different values of the speed at which the system is cyclically driven. The results are shown in 
Fig.\ref{fig:hist_phi_gz04}, with the qualitative features of the distribution closely resembling those obtained in the case with $\gamma_z = 0$.

Panel (a) of Fig.\ref{fig:hist_phi_gz04} corresponds to the faster case. The mean number of jumps $\bar{N}_J = 0.69$ is slightly above the one obtained in the $\gamma_z =0$. The additional jumps generated by $K_z$ are not sufficient to modify the distribution qualitatively, which continues to show a well-defined peak (arising from the occurrence of smooth evolution with no jumps) plus a broad small background. In panel (b), showing the case in which the system is driven slower, the mean number of jumps is also slightly increased from the $\gamma_z =0$ case due to the additional presence of $\gamma_z$ jumps, reaching a value $\bar{N}_J = 2.66$.  

The cases discussed above contain the first message of the present work. The stochastic nature of the GP in monitored dynamics needs to be taken into account and it is not possible to characterize it only through a single value. This raises the additional question of how this fact reflects on the experimental outcomes. To address this question, we will consider in the next Section a spin-echo protocol and see how, when, and whether the distribution in the interference fringes is affected by the randomness of the process. 

\subsection{Distribution of interference fringes in a spin-echo protocol} 
\label{sec:unrav}

If the system is prepared in an eigenstate of the Hamiltonian and subsequently driven in a cycle, adiabatically and in absolute isolation from the environment, then the quantum state accumulates a Berry phase that can be measured by implementing a spin-echo protocol~\cite{hahn}. It goes as follows. The system is initially prepared in a superposition state $\ket{\psi(0)}$ which reads $(1/\sqrt{2}) (\ket{\psi_+(0)} + \ket{\psi_-(0)})$ in terms of the ground and exited instantaneous eigenstates of $H(0)$. Then, it is driven for a period $T$, causing each eigenstate to acquire both a dynamical and a geometric phase $\phi_{\mbox{{\small a}}}^{\pm}$. A spin-flip operation and a second cycle in the opposite direction lead to a cancellation of the dynamical phases, resulting in a purely geometric relative phase. Berry phase can thus be extracted through state tomography~\cite{leek2007_cqed_observation, gasparinetti2016_cqed_observation, cucchietti} 
or by realizing that the probability for the system to be back in the initial state once the full evolution is completed, the persistence probability, is related to the Berry phase as $| \langle \psi(0) | \psi(2\,T) \rangle |^2 = \cos^2(2\,\phi^+_{\mathrm{a}})$ \cite{se_acotacion}.
The relation between the persistent probability and the GP given above relies on two factors: the adiabatic regime preventing the transitions between eigenstates and the exact cancellation of the dynamical phases during the protocol. If an echo experiment is performed on a system that is exposed to the effect of the environment and continuously  
monitored, the persistence probability will retain its dependence on the dynamical evolution. Nevertheless, it is worth understanding to which extent it is possible to learn features of GPs in a monitored system through an echo protocol.  

For each realization of the protocol, characterized by a sequence of jumps  $\mathcal{R}(2\,T, N_J)$, we can parametrize the persistent probability $\mathcal{P}_\mathcal{R}$ through an 
associated angle $\varphi_{\cal R}$ 

\begin{equation}
    \mathcal{P}_{\mathcal{R}} = | \langle \psi(0) | \psi(2T) \rangle |^2 \equiv \cos^2\left(2\, \varphi_{\cal R}\right).
    \label{eq:xdefinition}
\end{equation}
Both the persistence probability and the parameter $\varphi_{\cal R}$ inherit the stochastic character of the trajectories, with the probability of measuring a given value $\varphi$ related to the probability of the trajectories as

\begin{equation}
    P[\varphi] = \sum_{\mathcal{R}/\varphi_{\cal R}=\varphi} P[\mathcal{R}].
    \label{eq:vphase_probability}
\end{equation}
In the limiting case in which the persistence probability approaches its adiabatic value, $\varphi$ will approach $\phi^+_{\mathrm{a}}$. Away from that particular regime, $\varphi_{\mathcal{R}}$ is NOT equal to the GP $\phi_{\cal R} = \phi[\mathcal{R}]$ but, as mentioned previously, a convenient parametrization of the spin-echo interference fringes.

The non-adiabatic and environment-induced deviations from $\phi^{+}_\mathrm{a}$ can be analyzed by examining the ensemble $\{\varphi_\mathcal{R}\} =\varphi_{\{\mathcal{R}\}}$ that is obtained by computing Eq. (\ref{eq:xdefinition}) for each individual realization of the protocol. This study will also allow seeking possible relations, if any, between the stochastic behavior of the GPs and that of experimental outcomes (note that $\varphi_{\mathcal{R}}$ is defined modulo $\pi/2$ and up to a sign, therefore, any relation between the distribution of GPs and the distribution of the experimental results should take this into account). The frequency $\Omega$ at which the magnetic field is rotated is expected, once again, to have a direct impact on the distribution~\cite{sjoqvistshortcut, measuringshortcut}. On increasing the relative value of $\Omega$, the system will be exposed to the disruptive influence of the environment for shorter times, allowing to a larger extent a partial cancellation of the dynamical phases. At the same time, in this regime, non-negligible deviations from the adiabatic results will be unavoidable. On the other hand, smaller values of $\Omega$  might result in the system being exposed to environmental effects for too long, leading to strong deviations of the echo-parameter values $\varphi$ from $\phi^+_\mathrm{a}$. 

\begin{figure}[!ht]
    \centering
    \includegraphics[width = \columnwidth]{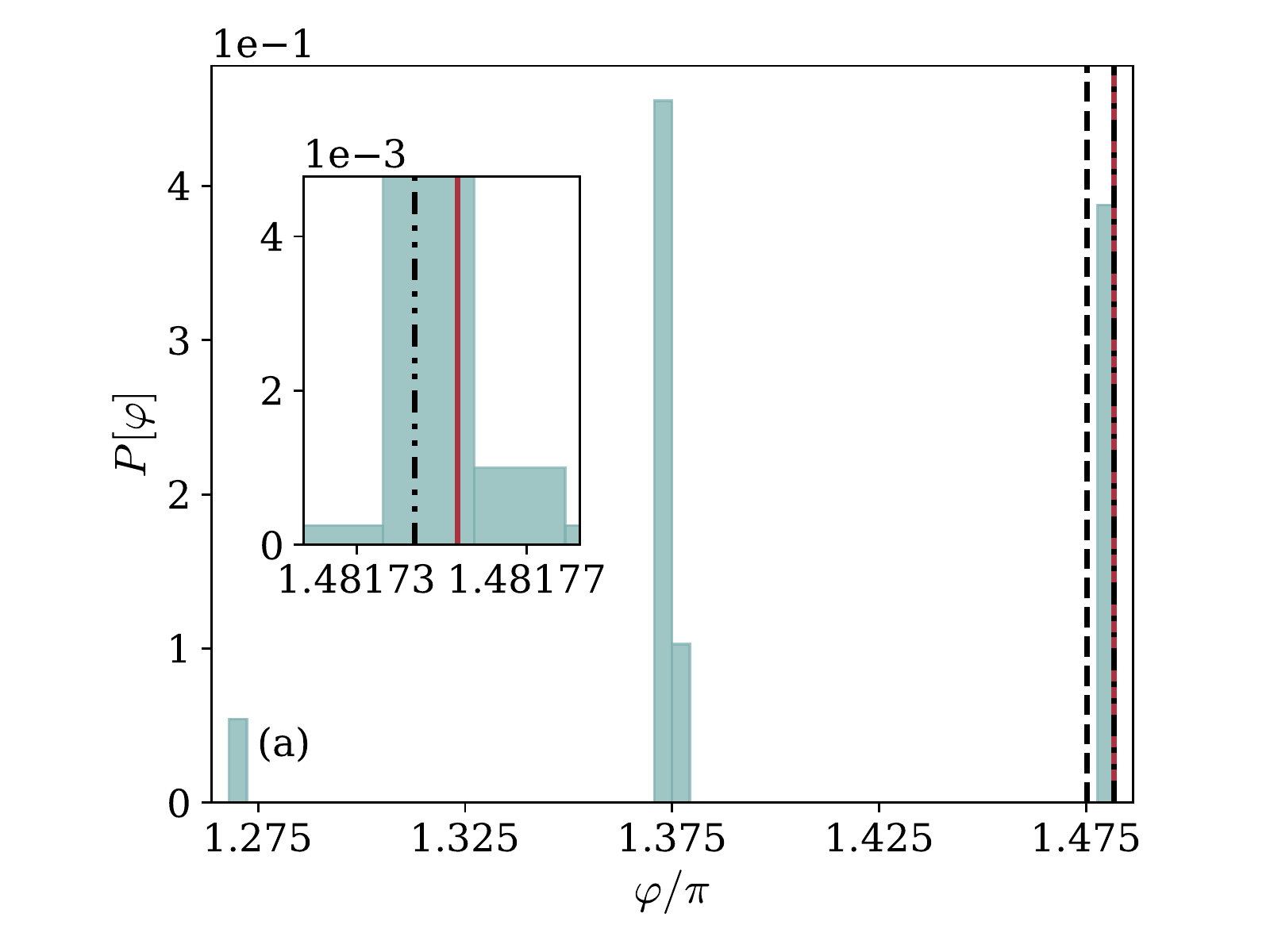}
    \includegraphics[width =  \columnwidth]{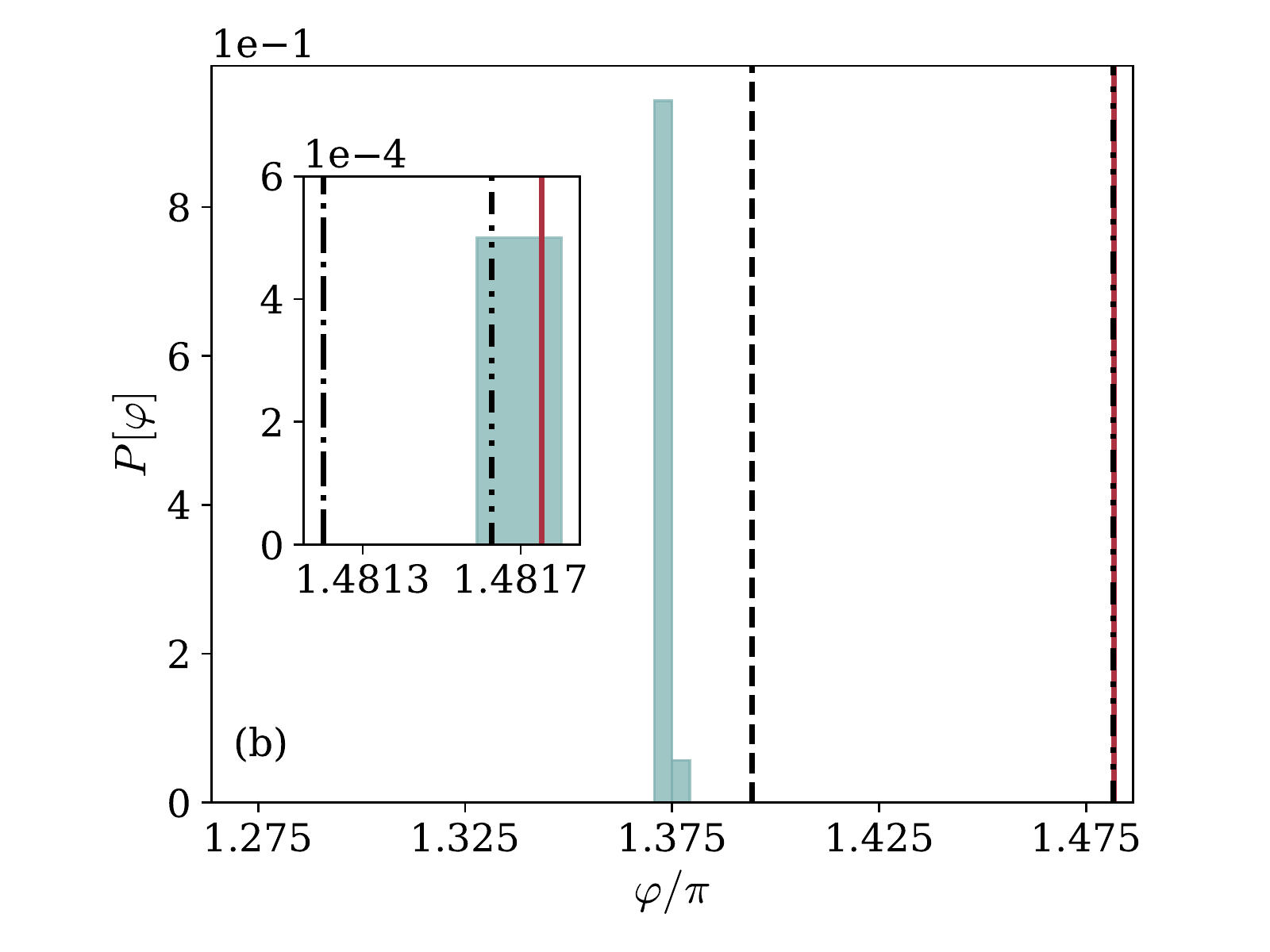}
    \caption{Probability distribution $P[\varphi]$ obtained in the echo-protocol for a magnetic field oriented with $\theta = 0.34\pi$ and driven in a loop at frequencies (a) $\Omega = 5\times 10^{-3}\omega$ and (b) $\Omega = 5\times 10^{-4}\omega$. The environment remains the same, characterized by the dissipation rate $\Gamma = 10^{-3}\omega$ and a $\gamma_z = 0$. In both panels, the solid red line depicts the adiabatic (Berry) phase $\phi^+_\mathrm{a}$, and the black dashed and dash-dotted lines signalize the GPs obtained in no-jump and unitary evolution respectively. 
    Furthermore, the black dash double-dotted line indicates the $\varphi$ value obtained in an echo protocol with no jumps. The insets in both panels show a range in which the result of a smoothly performed echo experiment is distinguishable from the Berry phase.}
    \label{fig:hist_varphi}
\end{figure}

In analogy with what we did in section \ref{sec:unrav_gp}, we examine first the case $\gamma_z=0$ and present, in Fig.\ref{fig:hist_varphi}, two representative cases in which the hypothetical {\em unitary} evolution would either be faster or slow enough to be considered within the adiabatic regime. These are shown in panels (a) and (b) of Fig.\ref{fig:hist_varphi} respectively. In both panels, we also display the adiabatic Berry phase $\phi_\mathrm{a}$ (which does not depend on $\Omega$), the GP $\phi_0$ obtained in a no-jump evolution, and the GP $\phi_u$ obtained in general unitary evolution. The $\varphi$ value obtained from an echo experiment which is completed without detecting jumps is also shown. For the parameters chosen, both panels show very small deviations of $\varphi$ extracted in a protocol with no jumps from the Berry phase (see  the insets in Fig.\ref{fig:hist_varphi}). It should be noted, however, that the probability of registering this specific trajectory is different in the two cases, as can be seen in the differences in the full $P[\varphi]$ distributions. 

The first striking feature that comes out is the presence of three distinct sharp peaks. The broad distribution observed in the GP values completely disappears in the spin-echo. 
This behavior originates from the fact that when $\gamma_z =0$, only jumps between instantaneous eigenstates are possible. This particular aspect of the unravelling leads, when combined with the properties of the persistence probability, to a distribution of interference fringes qualitatively different from that of the GPs. Each of the peaks shown in panel (a) of Fig. \ref{fig:hist_varphi} can be understood as arising from a different set of quantum trajectories in the following way. For the parameters chosen in panel (a) of Fig. \ref{fig:hist_varphi} trajectories with at most one jump are possible. The three peaks correspond to protocols with no jumps, protocols with one jump of the type $L_{\pm}$, and one jump of the type $L_d$ respectively. 
We refer to Appendix \ref{ap:echo_distribution} for a detailed justification of this identification. Trajectories that remain smooth along the whole protocol induce the right peak in Fig.\ref{fig:hist_varphi} (closest to the no-jump result, $\varphi\sim 1.475\pi$ for this choice of parameters). 
The central peak, centered at the value $\varphi\sim\,1.375\pi$ trivially associated via Eq. (\ref{eq:xdefinition}) with a persistence probability taking the value $1/2$, builds up from all those cases in which the state of the system is, at some given time, projected into an eigenstate of $H(t)$. In those trajectories, all the information about the accumulated phase before the jump is lost. As a consequence, immediately after a jump $L_{\pm}$, and regardless of both the previous evolution and the time at which the jump occurred, the persistence probability takes the exact value $1/2$. 
The third peak, the left one, is due to trajectories in which a jump $L_d$ occurs. This type of jump has the effect of introducing a $\pi$-shift in the relative phase of the echo state, that corresponds with the position of the left peak in Fig.\ref{fig:hist_varphi}. Therefore, the interference fringes distribution shows three peaks out of which two encode the same information, namely, the $\varphi$ value of a smoothly driven protocol, while the central peak contains almost no information.
Furthermore, the distribution is quite sharp because, for the parameters chosen, the described classes of trajectories are all detected, while more complex quantum trajectories are highly improbable (see Appendix \ref{ap:echo_distribution}). In panel (b) of Fig. \ref{fig:hist_varphi} the two peaks located at the sides have almost vanished. This reveals that when the system is driven at lower relative frequencies, a decay jump or a spontaneous excitation will be detected in almost every trajectory. A similar effect is obtained if the decay rate $\Gamma/\omega$ increases while keeping the ratio $\Omega/\omega$ fixed. 

A second aspect of the distribution $P[\varphi]$ is which features of GPs in open systems it captures. In panel (a) Fig. \ref{fig:hist_varphi}, the fast-driven regime, the $\varphi$ value obtained from protocols with no jumps agrees well with the adiabatic (Berry) phase, and both of these show small but visible deviations from no-jump GP. 
The $\varphi$ value is more closely related to the adiabatic case than the actual GP accumulated in smoothly drifted dynamics.
For the slower driving shown in panel (b) of Fig.\ref{fig:hist_varphi}, the no-jump $\varphi$ value remains a good indicator of the adiabatic phase, even though registering a smooth protocol is in this case less probable. Under these conditions, most of the experiment realizations will contribute to the central peak, which is not related to any characteristic GP. 

\begin{figure}[ht]
    \includegraphics[width = \linewidth]{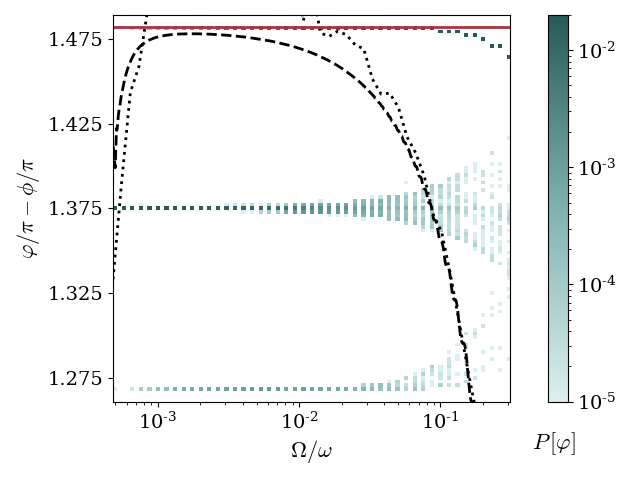}
    \caption{Probability distribution $P[\varphi]$ of $\varphi$ (as determined in an echo experiment) as a function of the ratio $\Omega/\omega$. The field is oriented with $\theta = 0.34\pi$ and the environment is characterized by the dissipation rate $\Gamma = 10^{-3}\omega$ and a $\gamma_z = 0$ . The $\varphi$ values are displayed on the y-axis, while the intensity of the count color indicates their probability. The solid red line depicts the adiabatic (Berry) phase $\phi^+_\mathrm{a}$, while the black dashed and dotted lines signalize the no-jump GP $\phi_0$ and the first moment $\Bar{\phi}$ of the GP distribution respectively.} \label{fig:varphi_vs_Omega}
\end{figure}

Inspection of Fig.\ref{fig:hist_varphi} suggests that, as in the case of the GPs distribution, the interplay between non-adiabatic corrections and environmentally induced jumps is better revealed when the distribution $P[\varphi]$ is analyzed as a function of the rate $\Omega/\omega$. This is shown in Fig. \ref{fig:varphi_vs_Omega}, which includes the two paradigmatic cases of Fig. \ref{fig:hist_varphi}. The Berry phase $\phi_{\rm a}$ and the values $\phi_{0}$ and $\Bar{\phi}$ of the GP associated with smooth trajectories and the first moment of the GP distribution are also given for reference. In the fast-driving regime, $\Omega/\omega \gtrsim 0.1$ the $\varphi$ value is most of the time the one arising in a protocol with no jumps and shows appreciable but still small deviations from the adiabatic phase. A trajectory with a single jump might be observed, albeit with less probability. If this is the case, the mixing of the eigenvalues due to non-adiabatic transitions will produce slightly broad distributions around the other two peaks, revealing the stochastic nature of the jump times. 
Non-adiabatic corrections have a much stronger impact on $\phi_0$ (for an analytical expression of this scaling, see Appendix \ref{ap:analytic}), its behavior completely disconnects from that of the distribution of echo protocols. 
On the other side, approaching the slow driving regime, the three peaks get sharper. This behavior is accompanied by a sharp decrease in the height of the side peaks and an enhancement of counts on the trivial, middle peak.  Along the full range, there is a region in which the interplay between environmentally-induced and non-adiabatic effects allows for good agreement between the GP accumulated in smooth non-unitary evolution and the value of $\varphi$. The behavior displayed by both the GP and the echo "phase" in smooth non-unitary evolution is further analyzed in Appendix \ref{ap:analytic}. 
The (consistently re-ranged) first moment of the GP distribution $\bar{\phi}$ remains, along the whole frequency range, uncorrelated from both the $\varphi$ distribution to a greater degree than all other characteristic values.

\begin{figure}[!ht]
    \centering
    \includegraphics[width = \columnwidth]{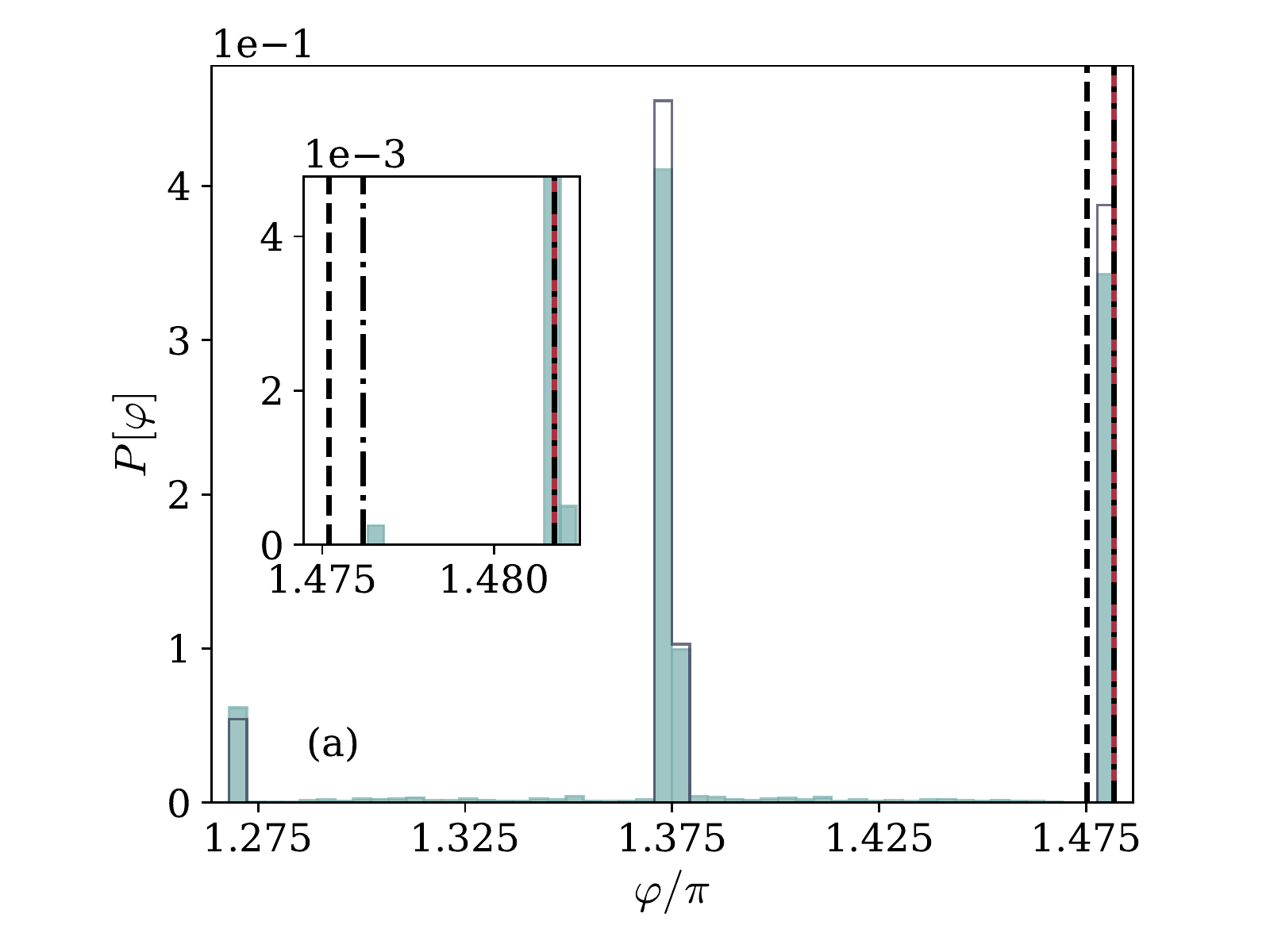}
    \includegraphics[width =  \columnwidth]{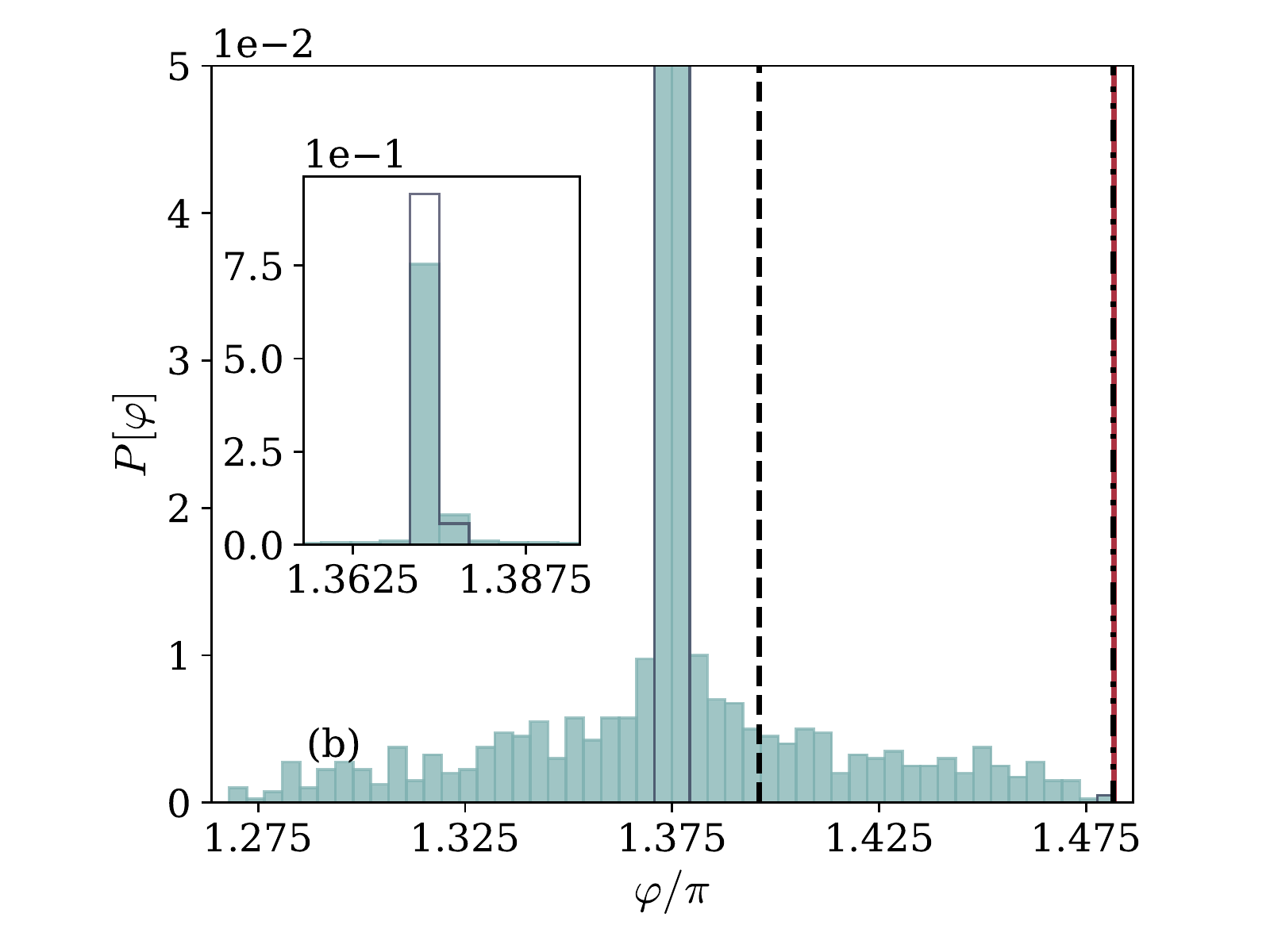}
    \caption{Probability distribution $P[\varphi]$ for a magnetic field oriented with $\theta = 0.34\pi$ and driven in a loop at frequencies (a) $\Omega = 5\times 10^{-3}\omega$ and (b) $\Omega = 5\times 10^{-4}\omega$. The environment is characterized by the dissipation rate $\Gamma = 10^{-3}\omega$, and finite $\gamma_z = 0.1\,\Gamma$. In both panels, a blue solid contour indicates the $\gamma_z = 0$ distributions. The solid red line depicts the adiabatic (Berry) phase $\phi^+_\mathrm{a}$, and the black dashed and dash-dotted lines signalize the GPs obtained in no-jumps and unitary evolution respectively. Finally, the black dash double-dotted line indicates the $\varphi$ value obtained in an echo protocol with no jumps. The insets zoom in a range in which differences between the reference values, panel (a), and the full magnitude of the central peak, panel (b), are visible.} 
    \label{fig:hist_varphi_gz}
\end{figure}

The distribution changes radically when $\gamma_z \ne 0$. In what follows we discuss the case $\gamma_z = 0.1\,\Gamma$ with $\Gamma = 10^{-3}\omega$. We start re-considering 
the two representative cases of fast and slower driving, displayed in panels (a) and (b) of Fig. \ref{fig:hist_varphi_gz} respectively. The first noticeable aspect is that, while three peaks observed in Fig. \ref{fig:hist_varphi} (indicated here by the blue contours) can still be detected, they are now coexisting with a broad distribution.  

\begin{figure}[ht!]
    \includegraphics[width = \linewidth]{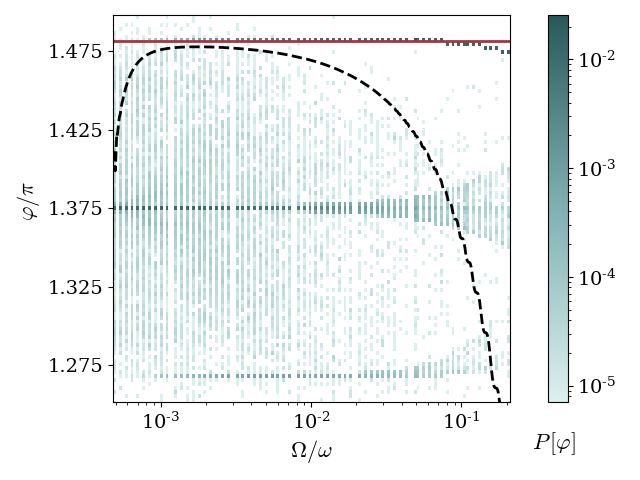}
    \caption{Probability distribution $P[\varphi]$ as a function of the rate $\Omega/\omega$ between the frequency $\Omega$ at which the magnetic field its rotated and its amplitude $\omega$. The field is oriented with $\theta = 0.34\pi$ and the environment is characterized by the dissipation rate $\Gamma = 10^{-3}\omega$ and $\gamma_z = 0.1 \Gamma$. 
    The $\varphi$ values are displayed on the y-axis, while the intensity of the count color indicates their probability. Extra lines signalize reference GP factors. The solid red line indicates the adiabatic (Berry) phase $\phi^+_\mathrm{a}$, while the black dashed line is the value $\phi_0$ extracted from evolution with no jumps.} 
    \label{fig:varphi_vs_Omega_gz}
\end{figure}

As visible in panel (a) of Fig. \ref{fig:hist_varphi_gz}, the three peak heights discussed previously decrease in the presence of $\gamma_z$. The suppression of the peaks is accompanied by the appearance of a broad background distribution covering the entire range. Panel (b) of Fig. \ref{fig:hist_varphi_gz} attends the slow driving situation, in which the probability to have a jump, and even several, along each trajectory, grows. The inclusion of the $L_z$ jump modifies the  sharp-peaked distribution into a broad one, which covers the entire range of $\varphi$ values. In particular, the two peaks connected to the no-jump trajectory disappeared. The inclusion of this term in the Lindbladian induces jumps into states other than the eigenstates of the Hamiltonian. In this sense, we may consider the results quite generic, not specifically dependent on the choice of the Lindblad operator. In order to get a more complete view of the effect of a finite $\gamma_z$, Fig. \ref{fig:varphi_vs_Omega_gz} shows the distribution of $\varphi$-values as a function of 
$\Omega/\omega$. 
For a fast-driven evolution in which almost no jumps are detected, the behavior exhibited by the distribution is similar to that observed in the $\gamma_z = 0$ case. 
When the velocity of the driving is reduced, gradually favouring the occurrence of jumps, the effect of introducing a finite  $\gamma_z$ value becomes more relevant. The $L_z$ jumps lead to $\varphi$ values that do depend on the time at which the jump took place and hence biuld up a broad background. 

Summarizing, while the distribution of interference fringes is, in general, quite different from that of the phase accumulated along a single trajectory, the analysis of a spin-echo protocol allows to extract reliable information on both the no-jump trajectories and the adiabatic (Berry) phase in some regimes of parameters. In the following Section we will concentrate on the no-jump trajectory (the kind of smooth evolution associated to the side-peaks of the probability distribution emerging in the echo-protocol) and show that undergoes a topological transition as a function of the coupling to the 
environment. 

%===================================================
%=================== SECTION =======================
%===================================================
\subsection{Topological transitions}\label{sec:topological}

As already anticipated, we conclude this analysis of GPs in monitored systems by focusing on the no-jump trajectory. We will show, following in spirit the work in Ref.~\cite{gefenWeak}, that the drift jump-free dynamics encode a topological transition. We would like to emphasize that, although the setting is very much different from that of ~\cite{gefenWeak}, we believe that the nature of the transition is the same. Our analysis is a strong hint to the conjecture that this type of transition is rather generic for monitored systems.

\begin{figure}[ht!]
    \includegraphics[width = \linewidth]{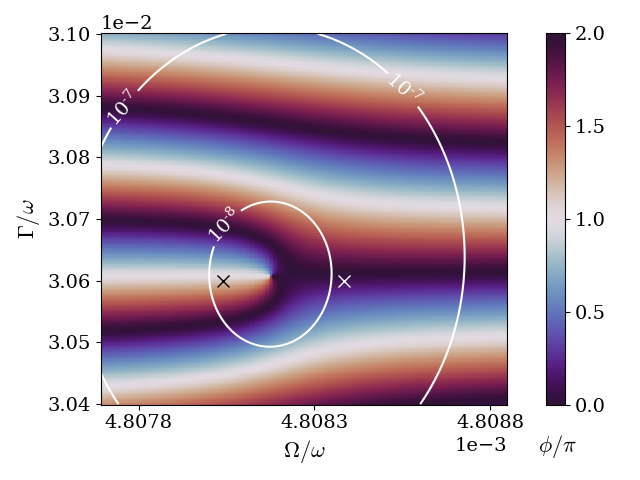}
    \caption{Geometric phase associated with the no-jump trajectory, displayed over a limited region of the parameters plane defined by the ratios  $\Omega/\omega$ and $\Gamma/\omega$. 
    The value of the GP is given by color, as indicated by the bar on the right. The direction of the field is fixed to $\theta = 0.34\pi$. A singularity is observed  $\Omega/\omega = 4.8082 \times 10^{-3}$ and $\Gamma/\omega = 0.0306$. The crosses indicate points slightly to the left of the singularity ($\Omega/\omega  = 4.8 \times 10^{-3}$) and slightly to the right of it 
    ($\Omega/\omega = 4.8084 \times 10^{-3}$), which will be shown to belong to different topological sectors.} 
    \label{fig:Colorplot}
\end{figure}

{\em Phase diagram - } The GP $\phi_0$ given by Eq. (\ref{eq:GP_nj}) depends, for every fixed $\theta$, on the ratios $\Omega/\omega$ and $\Gamma/\omega$. We recall the no-jump trajectory, and therefore the GP associated with it, have no dependence on $\gamma_z$. Plotted as a function of the above-mentioned parameters, the GP shows discrete singularities at critical points, around which it makes a $2\pi$ winding. Meanwhile, the probability associated with this particular trajectory vanishes at these points. We stress that, being the initial state of the system the eigenstate $\ket{\psi_+(0)}$ of $H(0)$, the null probability condition $\bra{\psi(0)}\ket{\psi(T)} = 0$ implies a final state $\ket{\psi_-(T)}$, meaning a singular point if found when full population transfer is attained within a time-interval $t\in[0,T]$, and refer to Appendix \ref{ap:analytic} for details on the analytical derivation. Fig.\ref{fig:Colorplot} shows a color plot of the GP in the $\Gamma-\Omega$ diagram at fixed values of the angle $\theta$. The range of the parameters is shown to highlight the singular point and the $2\pi$ winding of the GP around it. The white lines indicate the probability for the no-jump trajectory, which approaches zero on reaching the singularity.
We will show that the collection of these singular points delimits regions of the parameter space associated with different topological classes of evolution. This will be done by defining a topological invariant $\mathrm{n} \in \mathbb{Z}$ (see below) and explicitly showing it takes different values over different regions of the parameter-rates plane.

\begin{figure}[ht!]
    \includegraphics[width = \linewidth]{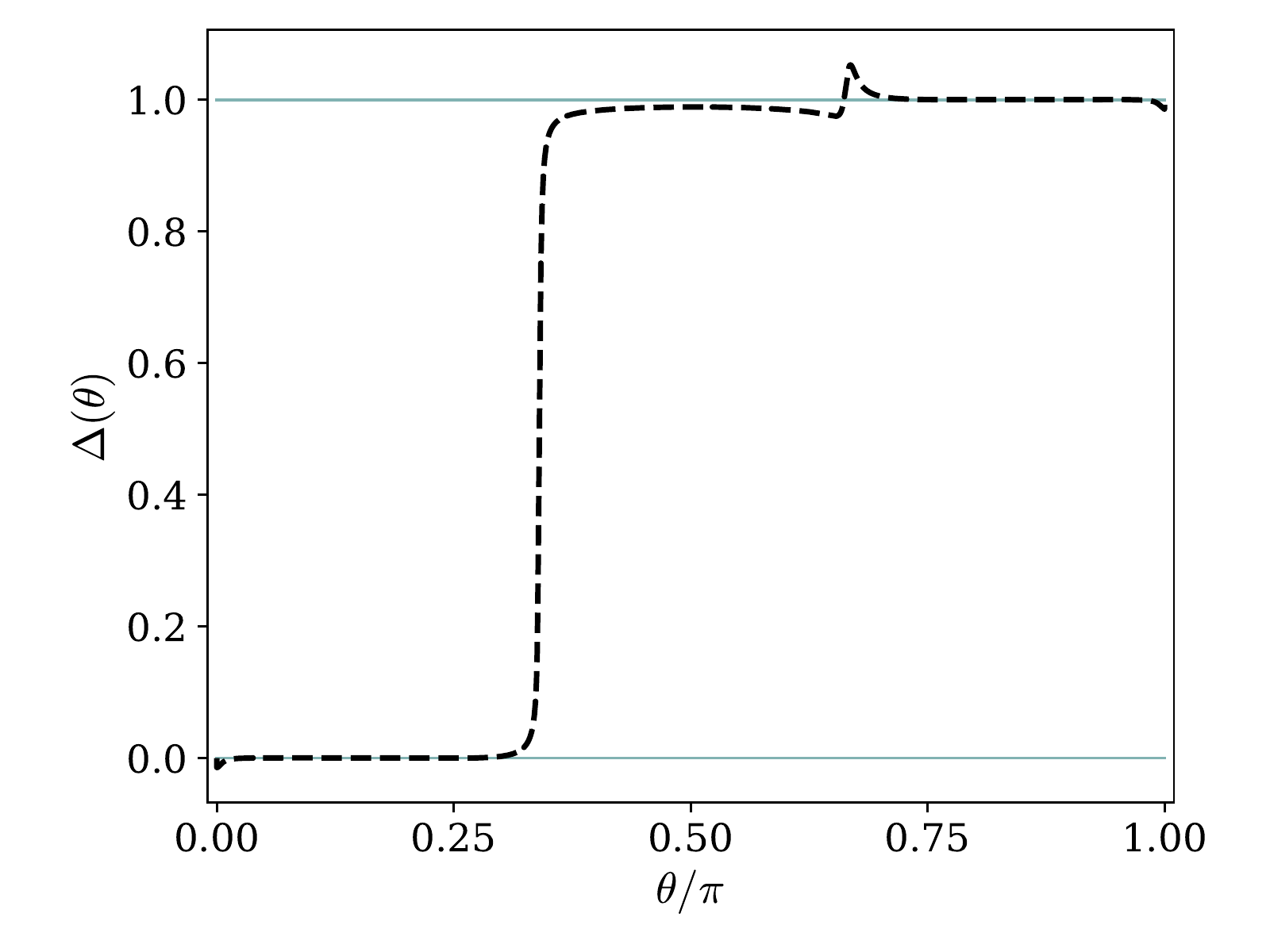}
    \caption{ $\Delta(\theta)$ between GPs computed for points slightly to the right and slightly to the left of the singularity, indicated in Fig. \ref{fig:Colorplot} with x's. The GP is, in each case, 
    characterized by a different value of the topological invariant $\mathrm{n}$. This is reflected in the fact that they differ by  $2\pi$ for $\theta = \pi$. \label{fig:Phi_resta}}
\end{figure}

{\em Topological transition in the no-jump trajectory - }
Direct inspection of the effective drift Hamiltonian shows that, if the magnetic field points in the z-direction, the exited eigenstate $\ket{\psi_+}$ of $H(t)$ remains fixed in a pole of the Bloch sphere independently of the values taken by the parameter rates $\Omega/\omega$ and $\Gamma/\omega$. Therefore, the GP associated with the no-jump trajectory identically vanishes (mod $2\pi$) for $\theta = 0$ and $\theta = \pi$. Without loss of generality, the mod $2\pi$ freedom can be eliminated from the GP by simultaneously setting $\phi_{0}(\theta = 0) =0$ and demanding continuity. 
In this way, $\phi_{0}(\theta=\pi)$ is completely determined by the evolution and acquires a value 

\begin{equation}
    \phi_{0}(\theta = \pi) = 2\pi\,\mathrm{n},
    \label{eq:n_definition}
\end{equation}
where $\mathrm{n}$ is an integer number that characterizes the dependence of the GP with $\theta$ for fixed parameter values. Being an integer, $\mathrm{n}$ constitutes a topological invariant because it can not be changed by smoothly deforming $\phi_0(\theta)$. As a consequence, if the GP is characterized by different values of $\mathrm{n}$ as a function of the various parameters, this will impose the GP to undergo a non-smooth transformation, as the singular behavior exhibited in Fig.\ref{fig:Colorplot}. 
Indeed, points in the parameter space slightly to the right and slightly to the left of the singularity (indicated with crosses in Fig. \ref{fig:Colorplot}) give rise to no-jump evolutions associated with topological invariants $\mathrm{n} = 0$ and $\mathrm{n} = 1$ respectively, thus identifying different topological classes. To explicitly show this, Fig. \ref{fig:Phi_resta} compares the behavior as a function of $\theta$ of these GPs by means of showing the difference $\Delta(\theta)$ between them. Given two points, say (1) and (2) and labelled by crosses in Fig. \ref{fig:Phi_resta}, 
$\Delta(\theta)$ is defined as 

\begin{equation} 
    \Delta(\theta) = \frac{1}{2\pi} \left[ \phi^{(\Gamma_1,\Omega_1)}_0 -  \phi^{(\Gamma_2,\Omega_2)}_0 \right].
    \label{eq:phi0_diff}
\end{equation}
This difference is seen to vanish (up to some smooth small deviations) up to $\theta = 0.34 \pi$, this is, until the angle of the singularity. At this specific $\theta$ value the GP obtained from each parameter rate abruptly deviates, so that their difference shows a step and settles around $\Delta = 1$ for the remaining range. The different topological numbers $\mathrm{n}$ is reflected by the value $\Delta(\pi) =1$ for $\theta = \pi$. 

\vspace{.2cm}

\begin{figure}[ht!]
    \includegraphics[width = \linewidth]{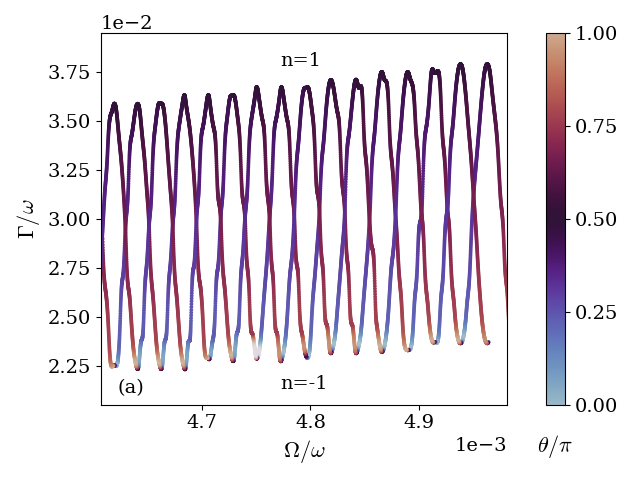}
    \includegraphics[width = \linewidth]{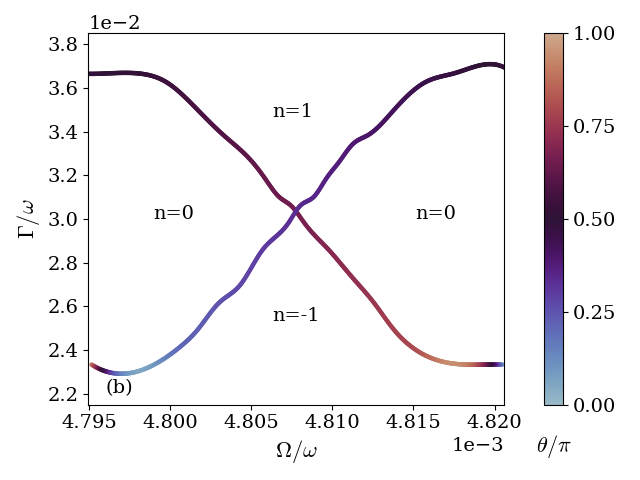}
    \caption{Critical lines dividing the parameters' plane into different topological classes of the no-jump evolution. The classes are characterized by different n values. The critical angle $\theta_c$ at which each singular point is found is indicated by a color as described by the bar on the right. Panels (a) and (b) display different ranges for the rates $\Omega/\omega$ and $\Gamma/\omega$.} 
    \label{fig:Singularities}
\end{figure}

\begin{figure}[!ht]
    \centering 
    \includegraphics[width = .95\columnwidth]{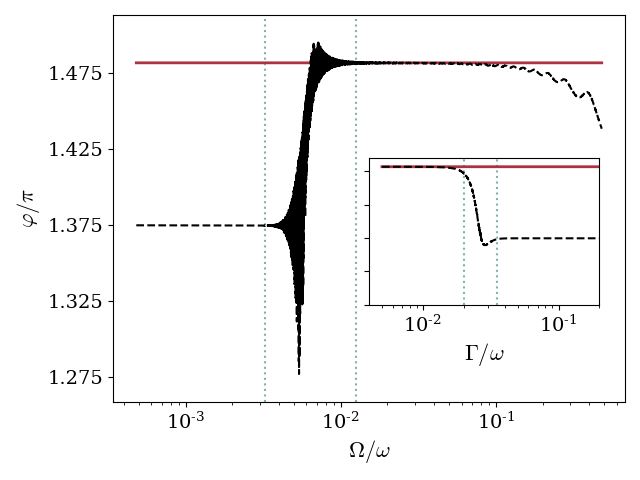}
    \caption{Dependence of $\varphi$ (black dashed line) obtained in a protocol with no jump events, as a function of the ratio $\Omega/\omega$. The field is oriented with $\theta = 0.34\pi$ and the environment is characterized by the dissipation rate $\Gamma = 0.0306\omega$, included in the ranges displayed in Figs. \ref{fig:Colorplot} - \ref{fig:Singularities}. The adiabatic (Berry) phase is also indicated for reference, with a red solid line. The inset shows the $\varphi$ value as a function of the rate $\Gamma/\omega$, with the magnetic field characterized by the same angle $\theta$ and $\Omega/\omega = 4.8\times 10^{-3}$, coinciding as well with the values used in the previous plots. } 
    \label{fig:varphi_deviation}
\end{figure}

Over the full parameter space, the GP shows several singularities, with locations that depend on the value of $\theta$. The set of singular points composes two counter-phase oscillating curves that define a chain of concatenated closed regions and split the parameter-rate space into an upper and lower region. This is shown in panel (a) of Fig. \ref{fig:Singularities}.
Parameters within each sector lead to the same $\mathrm{n}$ value. The area below the sequence of closed regions is characterized by $\mathrm{n}=-1$. The points given by parameter values $\Gamma = 0$ and $\Omega/\omega \ll 1$, defining the adiabatic regime, belong to this region. The  regions in between the lines are topologically trivial sectors with $\mathrm{n}=0$, while the upper one is characterized by $\mathrm{n} = 1$.  It is worth pointing out that these topological sectors are not equally probable. Besides the singular points of vanishing probability, 
the probability of attaining a trajectory with no jumps increases as $\Gamma$ is reduced. This implies that the upper topological sector is less probable than the others.  
\\
\\
{\em Topological transition in the echo experiment - } With the aim of seeking experimentally detectable signatures of the topological transition, we perform a close inspection of the echo experiment that is completed without any jump event. In Section \ref{sec:unrav}, the $\varphi$ value extracted in this case was observed to show good agreement with the adiabatic (Berry) phase for a wide range of frequencies. However, the close agreement of $\varphi$ with $\phi_{\rm a}$ will not hold for arbitrarily small frequency values, and it will deviate when the ratio $\Gamma/\Omega$ 
becomes sufficiently large. Fig. \ref{fig:varphi_deviation} shows the $\varphi$ value as a function of the frequency ratio. For easy reference and comparison, we consider an environment characterized by the dissipation rate  $\Gamma/\omega = 0.0306$, which is included in the ranges exhibited by Figs. \ref{fig:Colorplot} to \ref{fig:Singularities}. 

For a large frequency ratio, the no-jump $\varphi$ value shows the behavior described in Section \ref{sec:unrav}. However, approaching smaller frequencies, it shows a highly oscillating step and finally settles in the constant value $\varphi \sim 1.375\pi$, associated with a persistence probability $1/2$. Such a persistence probability is obtained when the state at the end of the protocol coincides, up to a global phase, with $\ket{\psi_-(0)}$, what happens when the smooth drift suppresses the occupancy of the exited eigenstate within a cycle leading to $\ket{\psi(T)}\sim \ket{\psi_-(T)}$.
As discussed, the parameter rates leading to full population transfer from the excited to the ground eigenstate exactly in a cycle of magnetic field rotation $t \in [0, T]$ correspond to the singular points observed in the phase diagram. Thus, full population transfer {\em within} the cycle requires the system to be driven at a slower frequency than that leading to a singular point, allowing therefore for greater exposition. This requirement establishes a connection between the value of the echo phase and the topological classes of evolution, as distinctive regimes of $\varphi$ are accessed on one and the other side of the singular points. We refer to Appendix \ref{ap:analytic} for details on this point. 
The limits of the range along which $\varphi$ shows the step and turns from $\sim \phi_{\rm a}$ into the central value are marked, on Fig. \ref{fig:varphi_deviation} with two light blue dotted lines. The righter region of the plot corresponds to evolutions characterized by the topological number $\mathrm{n} = -1$. The range between the light-blue lines corresponds to the densely packed sequence of topological sectors illustrated by panel (a) of Fig. \ref{fig:Singularities}. Finally, once on the left of the last vertical line, the evolution is associated with a 
value $\mathrm{n} = 1$ of the topological number.

The inset in Fig. \ref{fig:varphi_deviation} shows $\varphi$ as a function of the dissipation rate. In this plot, for easy reference and comparison, the value of the frequency rate is kept fixed at $\Omega/\omega = 4.8\times 10^{-3}$, also included in the ranges exhibited by Figs. \ref{fig:Colorplot} to \ref{fig:Singularities}. Once again, the $\varphi$ value shows good agreement with the adiabatic phase up to some critical $\Gamma/\Omega$ relation, at which it shows a decreasing step, finally landing at $\varphi \sim 1.375\pi$. As in the main plot, light blue dotted lines mark the limit of the step and split the plot into three distinctive sectors. The left of the first line corresponds to $\mathrm{n} = -1$ evolution, while the right side 
of the plot, to $\mathrm{n} = 1$. The space between lines, once again, can be associated with the intermediate zone, which is a single region (see Fig. \ref{fig:Singularities} (a)) thus leading to no oscillations.

In summary, a measure of the persistent probability in an echo protocol carries clear indications of the topological transition. The peak structure discussed in Section~\ref{sec:unrav} allows to identify the no-jump trajectory. The subsequent analysis of this peak, as summarized in Fig.\ref{fig:varphi_deviation}, is sufficient to capture the topological transition.

%=====================================================
%===================== SECTION =======================
%=====================================================

\section{Conclusions}\label{sec:concl}
In this paper, we have studied geometric phases in a continuously monitored quantum system. In absence of any coupling to the environment, the cyclic time-dependence of the Hamiltonian leads, in the adiabatic regime, to the Berry phase, and to its consistent generalization for a generic unitary evolution. The presence of an environment induces quantum jumps so that in a single realization of the dynamics the wave function, following a given quantum trajectory, accumulates a GP that is itself a stochastic quantity. We have analyzed the distribution of GPs by highlighting the interplay between non-adiabatic effects and the influence of the external environment. We have shown that for slow drivings the distribution of phases is broad because of the several different occurrences of jumps at random times. On speeding up the driving, the number of jumps reduces and the distribution becomes peaked around the no-jump trajectory (still deviating from the Berry phase because of the non-adiabatic correction and the non-Hermitian drift term). A first quantitative measure of the distribution has been given by the variance, discussed in Fig.\ref{fig:variance}. 

In order to have experimental access to the GPs along a given trajectory, we have also analyzed a spin-echo protocol. The structure provided by the jump operators taken together with the possibility of level transitions due to non-adiabaticity and the characteristics of the persistence probability can be set in such a way that they lead either to the observation of broad distributions or extremely sharp peaks. This interplay should be thus considered in order to be explored as a tool or otherwise, the experiment is rendered uninformative.

We have finally concentrated on the no-jump trajectory, showing that it undergoes a topological transition as a function of the dissipation strength. Interestingly, this transition is not necessarily connected to singularities occurring in the dynamics of the density matrix. Indeed, for the model considered herein, at the transition point occurring in the no-jump trajectory the behavior of the 
density matrix is smooth. Despite the striking differences shown between the GP and the interference fringes of an echo experiment, traces of this transition can be observed in the behavior of the interference fringes.

In this work, we have considered a specific model for the jump operators corresponding to a well-defined type of monitoring. However, it is important to understand to which extent the properties we have discussed here depend on the type of unravelling. This question might be of particular relevance, especially if one wants to define topological properties associated with Markovian systems starting from the properties of their trajectories (there are infinite ways of unravelling the same Lindblad dynamics). A glimpse on this question is summarised in Appendix~\ref{ap:diff_unrav} where we consider an unraveling corresponding to a homodyne detection. For what concerns the distribution the qualitative pictures we have outlined in the body of the paper remain valid although important quantitative differences may arise.

\begin{acknowledgments}
We would like to acknowledge Alessandro Romito for very useful discussions and critical reading of the manuscript. The work  of R.F. has been supported by the ERC under grant  agreement n.101053159 (RAVE) and by a Google Quantum Research Award. The work of L.V., F.L., and P.I.V. is supported by Agencia Nacional de Promoci\'on Cient\'\i fica y Tecnol\'ogica (ANPCyT), Consejo Nacional de Investigaciones Cient\'\i ficas y T\'ecnicas (CONICET), and Universidad de Buenos Aires (UBA). P.I.V. acknowledges ICTP-Trieste Associate Program. R.F. acknowledges that his research has been conducted within the framework of the Trieste Institute for Theoretical Quantum Technologies (TQT).
\end{acknowledgments}

\appendix
\section{Pancharatnam phase along a quantum trajectory}
\label{ap:gp_derivation}

As stated in section \ref{sec:th_qtraj}, the quantum trajectory emerging in a single monitored evolution of the system can be understood as intervals of smooth dynamics interrupted at random times by quantum jumps. Considered in this way, evolution in a time interval $t \in [0, T]$ is characterized by an array of jumps of type $\alpha_i$ occurring at times $t_i$ of the form given by Eq. (\ref{eq:R}), and the parameter $t$ is a continuous variable within the intervals delimited by the $t_i$'s. In the quantum jumps approach, the algorithm applied in constructing the trajectories goes as follows~\cite{Molmer:93}. The time interval $[0, T]$ is discretized into N steps of length $\delta \,t$. and the state is consistently updated at each time step according to a randomly-decided non-hermitian operator, as described in Eq. (\ref{eq:monitored-evol}). Hence, each quantum trajectory can also be thought of from an algorithmic point of view as the ordered collection of states generated by the action of a specific sequence of operators $K_{0, \alpha}$ given by Eq. (\ref{eq:jumpoperators}), and is in this way a discrete set of states.

For a sequence of N discrete pure states, the suitable GP expression is Pancharatnam phase~\cite{Carollo_original,Carollo_review, mukunda93}, and is given by

\begin{equation}
    \phi_P[\psi] = \arg\bra{\psi_1}\ket{\psi_\mathrm{N}}-\arg(\bra{\psi_1}\ket{\psi_{2}}...\bra{\psi_{\mathrm{N}-1}}\ket{\psi_{\mathrm{N}}}).
    \label{eq:phiDiscreta}
\end{equation}

The Pancharatnam phase is independent of the $U(1)$ gauge choice and does not require the sequence to close, rely on adiabaticity condition or demand for unitarity, allowing for non-normalized states in the sequence, as long as non of them perfectly vanishes. Exhibiting these characteristics, it becomes a natural definition of GP to be applied to monitored dynamics, in which evolution is generated by non-hermitian operators. It equals the unitary GP associated with the trajectory build-up from joining consecutive states in the sequence by the shortest geodesic in the Hilbert space.

\begin{figure}[ht!]
    \centering
    \begin{tikzpicture}
        \draw[thick]     (0.0,0) -- (0.6,0);
        \draw[thick]     (1.1,0) -- (5.2,0);
        \draw[thick, ->] (5.7,0) -- (7.0,0);
        
        \filldraw[black] (0.75,0) circle (.4pt);
        \filldraw[black] (0.85,0) circle (.4pt);
        \filldraw[black] (0.95,0) circle (.4pt);
        
        \filldraw[black] (5.35,0) circle (.4pt);
        \filldraw[black] (5.45,0) circle (.4pt);
        \filldraw[black] (5.55,0) circle (.4pt);
        
        \draw[thick]     (0.25,-0.2) -- (0.25, 0.2);
        \draw[thick]     (0.25,-0.2) -- (0.32,-0.2);
        \draw[thick]     (0.25, 0.2) -- (0.32, 0.2);
        \node at         (0.25,-0.5) {$0$};
        
        \draw[thick]     (6.5,-0.2) -- (6.5, 0.2);
        \draw[thick]     (6.5,-0.2) -- (6.43,-0.2);
        \draw[thick]     (6.5, 0.2) -- (6.43, 0.2);
        \node at         (6.55,-0.5) {$T$};
        
        \draw[thick]     (1.5,-0.15) -- (1.5,0.15);
        \node at         (1.5,-0.5) {$t_i$};
        \draw[thick, ->] (1.5, 0.25) -- (1.5,0.8);
        \node at         (1.5, 1.1) {$\ket{\psi(t_i)}$};
        
        \draw[thick]     (4.5,-0.15) -- (4.5,0.15);
        \node at         (4.5,-0.5) {$t_{i+1}$};
        
        \draw     (1.75,-0.1) -- (1.75,0.1);
        \draw     (2.00,-0.1) -- (2.00,0.1);
        \draw     (2.25,-0.1) -- (2.25,0.1);
        \draw     (2.50,-0.1) -- (2.50,0.1);
        \draw     (2.75,-0.1) -- (2.75,0.1);
        \draw     (3.00,-0.1) -- (3.00,0.1);
        \draw     (3.25,-0.1) -- (3.25,0.1);
        
        \draw[thick, ->] (3.5, 0.25) -- (3.5,0.8);
        \node at         (3.5, 1.1) {$\ket{\psi(t_i + k_i\,\delta t)}$};
        \node at         (3.5, 1.5) {\textcolor{white}{.}};
        \draw[thick, ->] (1.75,-0.5) arc (250:310:1.7);
        \node at         (2.5, -1.0) {$+ k_i\,\delta t$};
        
        \draw     (3.50,-0.1) -- (3.50,0.1);
        \draw     (3.75,-0.1) -- (3.75,0.1);
        \draw     (4.00,-0.1) -- (4.00,0.1);
        \draw     (4.25,-0.1) -- (4.25,0.1);
        
        \draw[gray] (0.50,-0.1) -- (0.50,0.1);
        \draw[gray] (1.25,-0.1) -- (1.25,0.1);

        \draw[gray] (4.50,-0.1) -- (4.50,0.1);
        \draw[gray] (4.75,-0.1) -- (4.75,0.1);
        \draw[gray] (5.00,-0.1) -- (5.00,0.1);
        
        \draw[gray] (6.00,-0.1) -- (6.00,0.1);
        \draw[gray] (6.25,-0.1) -- (6.25,0.1);

    \end{tikzpicture}
    \caption{Illustrative diagram depicting time interval $[0,T]$. Both the discretization in $\delta t$ steps and the splitting at jump times $t_i$ are indicated. The relation 
    between times and states is represented as well \label{fig:time_interval}}
\end{figure}
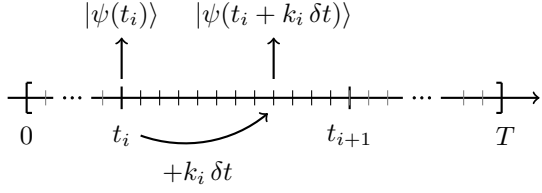

While this definition does not imply any constraint on the number of states in the sequence by itself, when applied in the context of quantum jumps the number N of states is constrained from below as a consequence of the condition reigning the time step. An evolution in time-interval $[0,T]$ consist of $\mathrm{N} =T / \delta\,t \gg1$ states. Splitting the sequence of states $\{\ket{\psi_1} \ket{\psi_2}... \ket{\psi_\mathrm{N}}\}$ into sets starting and ending at those corresponding to the specific times $t_i$ where a jump is registered, sets a bridge between this two different 
descriptions of a quantum trajectory. Each time interval $[t_i, t_{i+1}]$, discretized in time-steps of length $\delta t$, consist of a number of steps that depend on the specific values of $t_i$ and $t_{i+1}$. From a given jump-time $t_i$, any time-step in the consecutive interval can be found as $t_i + k_i\,\delta t$, this is, by adding some amount $k_i \in \mathbb{N}$ of increments $\delta t$, up to some maximum value $k_i^*$ that satisfies $t_{i+1} = t_i + k^{*}_i\,\delta t$ (See Fig. \ref{fig:time_interval}).

At each given time, the outcome of a measurement performed on the environment will be associated to the corresponding Kraus operator acting on the system and the state generated by its action. Therefore, there is a to a one-to-one correspondence between the discrete set conforming the time interval and the array of states forming the trajectory. The splitting at jump-times $t_i$ can thus be 
mapped into the trajectory as

\begin{equation}
    \bigcup_{i=0}^{N_J}\{\ket{\psi(t_i + k_i\,\delta t)} k_i = 0, ..., k^{max}_i -1\}
\end{equation}
with $N_J$ the number of jumps occurring in the trajectory and the out-bounds indexes $i=0$ and $i=N_J +1$ signaling the entire time-interval limits $t_{0} = 0$ and $t_{N_J +1} = T$.

Introducing such a decomposition into the formula for Pancharatnam phase, Eq. (\ref{eq:phiDiscreta}) can be re-written as

\begin{align}\nonumber
    \phi_P =& \arg\bra{\psi(0)}\ket{\psi(T)}\\\nonumber
    -& \sum_{i=0}^{N_J}\sum_{k_i=1}^{k^{*}_i-1} \arg\bra{\psi(t_i + k_i\,\delta t)}\ket{\psi(t_i + (k_i+1)\,\delta t)}\\
    -& \sum_{i=0}^{N_J} \arg\bra{\psi(t_i)}K_{\alpha_i}\ket{\psi(t_i)}.
\end{align}
The formula in Eq.(\ref{eq:GP_traj}) for the GP is thus associated with a single trajectory is derived by taking the continuous limit $\delta t/T\rightarrow 0$ within the intervals of smooth 
evolution~\cite{Carollo_original, Carollo_review}. This expression, more suitable for the exam performed in our work, inherits all the properties of the Pancharatnam phase from which it is obtained.

\section{Interference fringes distribution}
\label{ap:echo_distribution}
As discussed in Sec. \ref{sec:unrav}, the distribution of interference fringes from an echo experiment, which we parameterize with $\varphi$, shows three (sometimes sharp) peaks. 
When $\gamma_z =0$ only jumps between instantaneous energy eigenstates are possible, and the three peaks emerge from sets of trajectories of a different character as follows.

\begin{enumerate}
    \item Smooth protocols with no jumps generate the piling up in the no-jump value $\varphi_0 \sim 1.43\pi$
    \item Protocols in which at least one decay or spontaneous excitation jump occurred, projecting the state into an eigenstate $\ket{\psi_\pm(t_i)}$ of $H(t)$, give rise to the peak at $\varphi \sim 1.375\pi$.
    \item Protocols in which only dephasing jumps took place give rise to the peak at $\varphi\sim1.275\pi$.
\end{enumerate}

In this appendix, we provide a detailed justification of this observation. With the aim of providing an accessible presentation of the qualitative aspects of the phenomena, we will generally disregard the non-hermiticity of the smooth evolution between jumps, thinking of those intervals as unitary (slowly or rapidly driven) evolution. Hence, this presentation should not be taken as a rigorous quantitative analysis.

We begin with the consideration of the peak (1.) coinciding with the no-jump value $\varphi_0\sim 1.34\pi$. As presented in Sec. \ref{sec:th_qtraj} this smooth trajectory is unique and therefore the exact same value of $\varphi$ will be expected on every case in which this trajectory is obtained.

We thus turn to the case in which jumps are indeed detected, with special care on the anti-intuitive shrinking of the distribution in the slower regime in which more jumps are detected. 
When $\gamma_z = 0$ three jumps are possible within our unravelling of the Lindblad equation. Two out of these three jumps project the state into an energy eigenstate, namely, decay jumps and spontaneous excitations. Whenever a jump of this kind takes place at some instant of time $t_i$, immediately after the jump the state of the system turns into 
\begin{equation}
    \ket{\Psi(t_i)}= e^{i\,\xi(t_i) + i\,\phi(t_i)}\ket{\psi_\pm(t_i)}
\end{equation}
with $\xi(t_i)$ the dynamical phase and $\phi(t_i)$ the geometrical phase, given by Eq. (\ref{eq:GP_traj}) accumulated up to the occurrence of the jump. If the protocol ends immediately after, the persistence probability  $\mathcal{P}_\mathcal{R} = |\bra{\psi(0)}\ket{\psi(2T)}|^2 = 1/2$ preserves no information on either the GP or the specific characteristics of the jump. If, on the other hand, the system continues to evolve, the possibility of obtaining any information on a phase or the jump time will rely on the interplay between the non-adiabatic transitions and the existence of further jumps. 
If the evolution continues from the first jump on, this will happen smoothly until either the protocol is finished or another jump takes place. Different regimes of $\Omega/\omega$ give rise to the smooth evolution of different natures. If the protocol is performed slowly enough, this smooth evolution is (almost) transition-free and $\ket{\psi(t)}\sim e^{i\,\xi(t>t_i) + i\,\phi(t>t_i)}\ket{\psi_\pm(t>t_i)}$, 
so the result obtained for the persistent probability remains to be $\mathcal{P}_\mathcal{R} = 1/2$. Moreover, this regime favors the occurrence of further jumps, thus reinforcing the erasing of information by re-projecting into eigenstates of $H(t)$. The complete independence of the result on the times $t_i$ of the jumps makes this peak (2.) extremely sharp in the slow regime.
On the other hand, if the system is driven faster, along the smooth evolution after the jump the state develops contributions from the other eigenstate due to non-adiabatic effects, favoring the emergence of relative phases and becoming
\begin{align} \nonumber
    \ket{\psi(t>t_i)} = A_\pm(t>t_i) \ket{\psi_\pm(t>t_i)} \\+ A_\mp(t -t_i) \ket{\psi_\mp(t-t_i)}
\end{align}
with $A\pm(t)$ the amplitudes for each eigenstate. In such a situation, the persistence probability depends on $t_i$, leading to the broadening observed in the central peak of 
Fig.\ref{fig:varphi_vs_Omega} for faster driving, while still not trivially connected to the GP. As anticipated in the previous paragraphs, each jump of this kind will erase all information on the phases and any dependence on previous jump times. The possibility of further erasing events is mitigated in faster protocols by the reduction of exposure to the environment.  

The third peak (3.) observed in the distribution at $\varphi \sim 1.475\pi$ can be understood by adding dephasing jumps to the previous discussion. A dephasing jump has the effect of 
introducing a $\pi$ shift in the relative phase of the state. If the evolution afterward remains transition-less (and no erasing jumps occur at any point), the evolution resembles that of the adiabatic echo experiment up to corrections that can be disregarded, so the persistence probability takes the value $\mathcal{P}\sim\sin^2(2\phi_\mathrm{a})$ (with $\cos$ replaced by $\sin$ due to the relative $\pi$ shift). This situation leads to a well-defined single $\varphi$-value which is independent of the time $t_i$ at which the jump took place. Therefore, in the slow-driving range, a well-defined peak emerges, that might however be small, as in this regime decay jumps are likely. As the magnetic field is rotated faster, non-adiabatic effects induce a dependence on $t_i$ on the persistence probability. This dependence on $t_i$ is inherited by the "phases" extracted, and thus responsible for the broadening of the distribution observed in Fig. \ref{fig:varphi_vs_Omega} for larger $\Omega/\omega$ values. 

The inclusion of a jump operator $\propto \sigma_z$ modifies this three-peaked distribution by leading to a broad background which is present even in the case in which it is not the dominant process. The $K_z$ jumps promote the development of relative phases as they mix eigenstates of the Hamiltonian. Even if the system has, at some time, transitioned to an eigenstate, suffering from a $\sigma_z$-jump suddenly drags it away into a superposition state.

\section{Dependence on the unravelling: field displacement}
\label{ap:diff_unrav}

Another paradigmatic quantum trajectories scheme arising from a different unraveling of the master equation is that of the so-called diffusive trajectories, in which the monitored quantities produce continuously fluctuating signals instead of discontinuous jumps~\cite{acotacion}. This is the prototypical scheme of continuous or ideal homodyne detection, which can be theoretically obtained as a limiting case of the mentioned discrete homodyne detection~\cite{carmichael1993_open, wiseman, wiseman2009quantum, percival1998quantum, sjoqvist2010_hidden}.

The master equation Eq. (\ref{eq:Lindblad}) is invariant under the transformation

\begin{align}\nonumber
    &H(t) \rightarrow H'(t) = H(t) -\sqrt{\lambda}\frac{i}{2}\sum_\alpha(K_\alpha - K_\alpha^\dagger)\\
    &K_\alpha \rightarrow K'_\alpha = K_\alpha + \sqrt{\lambda}\,\mathbb{I},
    \label{eq:Transformation}
\end{align}
where $\sqrt{\lambda} \in \mathbb{R}$.
Therefore it is possible to substitute $K_\alpha$ and $H(t)$ in Eq. (\ref{eq:Lindblad}) by $K'_\alpha$ and $H'(t)$ without modifying the averaged dynamics of the system and unravel it using the standard direct detection (quantum jumps) scheme applied before. When the reservoir is assumed to be made of harmonic modes, like electromagnetic radiation, adding the 
displacement $\sqrt{\lambda}$ to the Lindblad operators corresponds to the implementation of homodyne detection~\cite{wiseman2009quantum, wiseman, sync}.
In this case, taking $\sqrt{\lambda}$ suitably large leads to a measurement of the quadrature of the system dipole $K_\alpha + K^\dagger_\alpha$. However, in order to keep the collapse 
probability per step small, it would be necessary to reduce the time step and hence increase the simulation time by the same order. For this reason, we refrain to consider finite large $\sqrt{\lambda}$ values in this section and focus on the modifications suffered by $P[\phi]$ for smaller $\sqrt{\lambda}$ values.

\begin{figure}[ht]
    \centering 
    \includegraphics[width = .95\columnwidth]{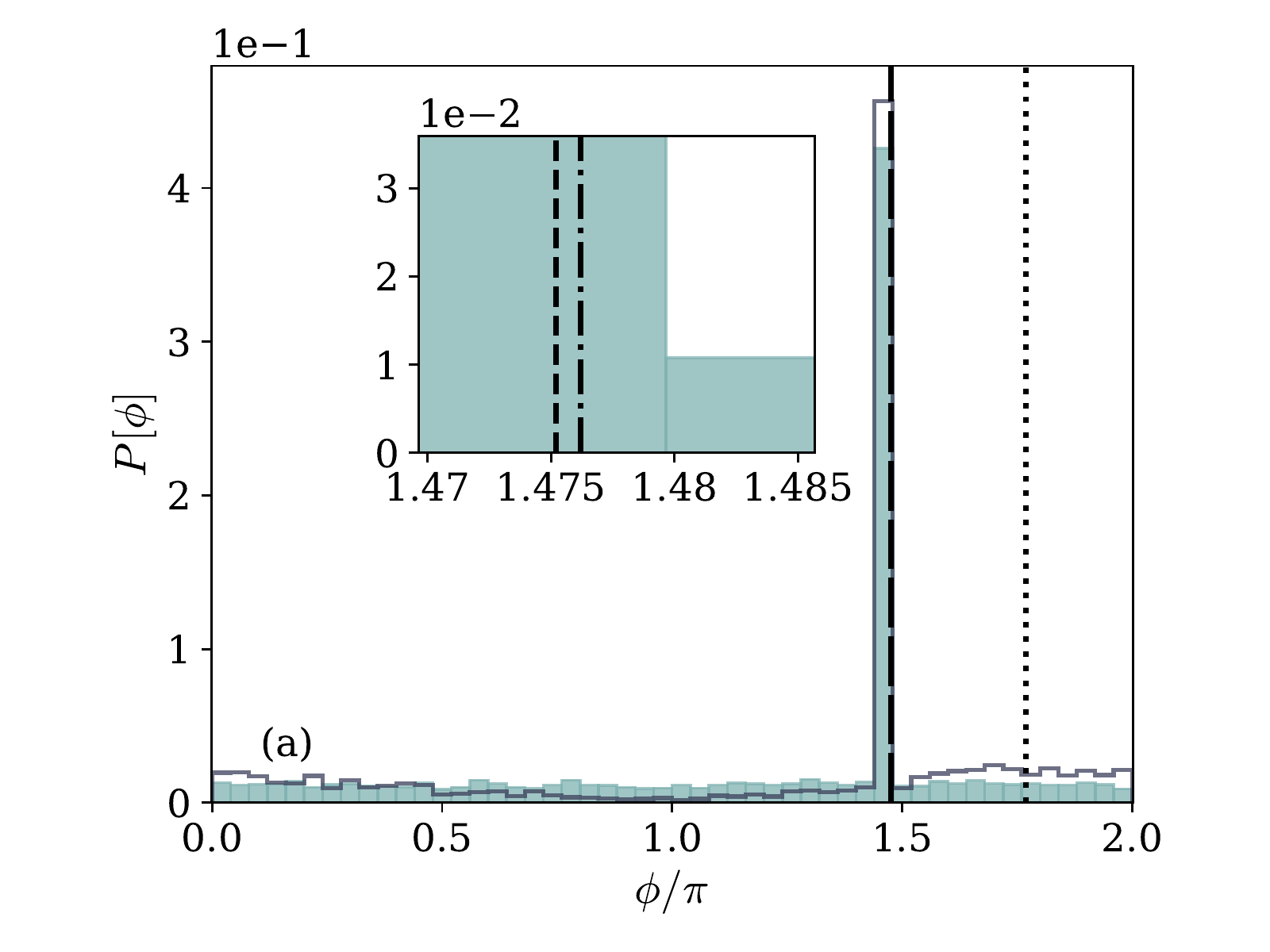}
    \includegraphics[width =\columnwidth]{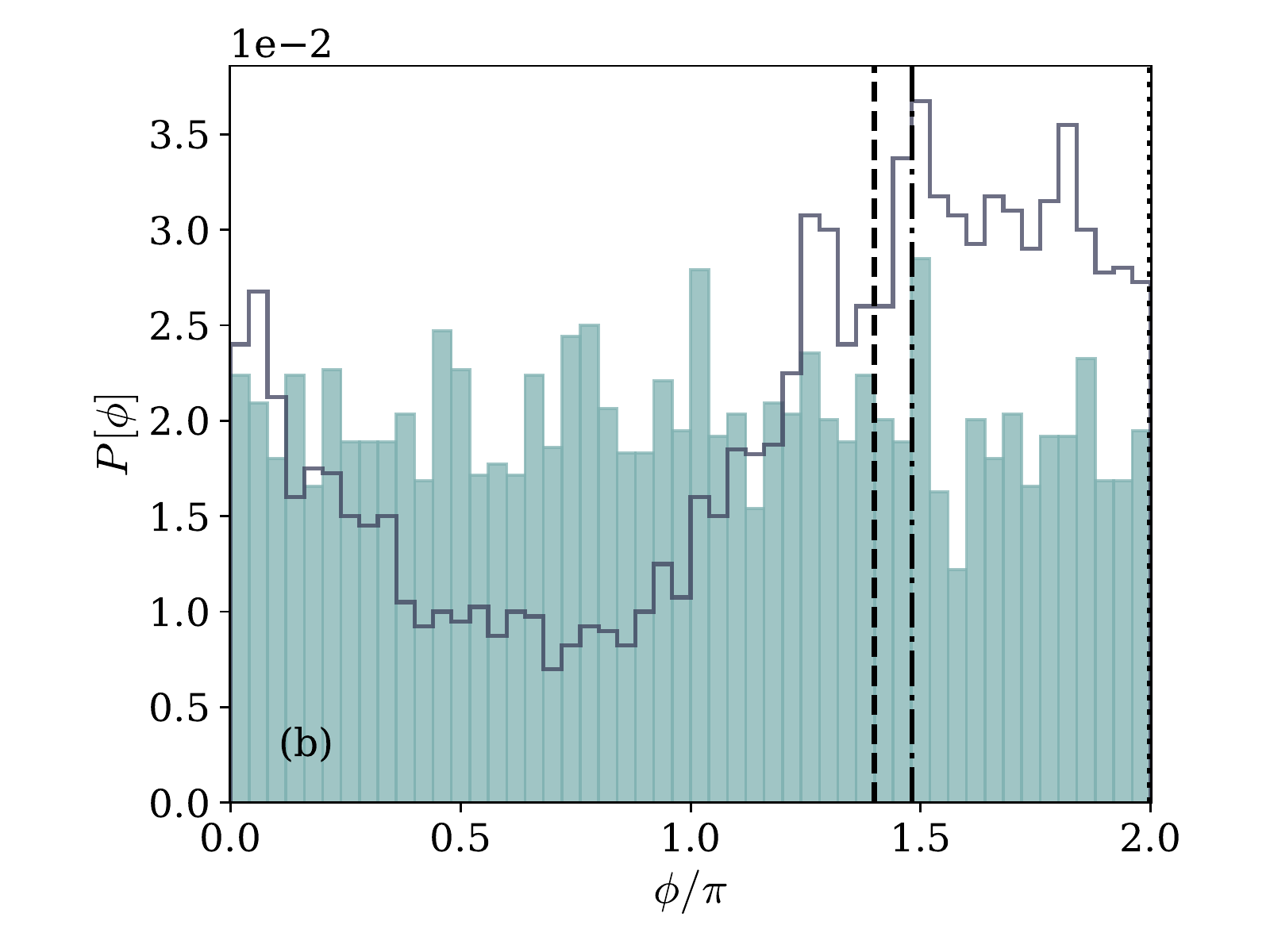}
    \caption{Probability distribution $P[\phi]$ of GP values obtained in an unravelling with $K'_\alpha$ and $H'(t)$ operators. The magnetic field is oriented at $\theta = 0.34\pi$ and driven in a cycle at frequencies (a) $\Omega = 5\times 10^{-3}\omega$ and (b) $\Omega = 5\times 10^{-4}\omega$. The environment is characterized by the dissipation rate $\Gamma = 10^{-3}\omega$ and $\gamma_z = 0$. We have taken $\lambda = 2.5\times 10^{-5}\omega$. In both panels, a blue solid contour recalls the distributions obtained in the original unraveling considered in this work. Extra lines indicate the new reference GP values. The black dashed and dot-dashed lines signalize the GPs $\phi_0$ and $\phi_u$ associated with no-jump and general unitary evolution. The black dotted line shows the first moment $\Bar{\phi}$.}  
    \label{fig:unravelings}
\end{figure}

In Fig. \ref{fig:unravelings} we present, also for the case of this different unravelling of the Lindblad equation, the two cases in which the driving is performed faster or slow enough for the hypothetical {\em unitary} dynamics to be considered adiabatic. As in the previous cases, the environment remains fixed with $\Gamma = 10^{-3}\omega$ and $\gamma_z = 0$, and we have taken $\lambda = 2.5\times 10^{-5}\omega < \Gamma$. The two cases are shown in panels (a) and (b) of Fig.\ref{fig:unravelings} respectively, where we also plot the no-jump and unitary GPs, and the average of the distribution for reference. 

Striking differences from the case of direct detection arise. For this $\lambda/ \omega\ll1$, the reference values displayed remain close to those obtained in the $\lambda = 0$, while 
the distributions behave differently. In the fast-driven case displayed in panel (a), the expected increase in jumps is reflected by the decrease of the sharp peak piling up from no-jump trajectories. However, the formerly broad, but still uneven background, has now turned into a completely uniform distribution in which each phase value (but the no-jump) is evenly probable. The described behavior is reinforced when the system is driven at slower frequency rates. 
The previously broad while single-peaked distribution lacks, in a system monitored through the operators $K'_\alpha$ forming the new basis, of any structure.

\section{Smooth evolution with no jumps: analytic approach\label{ap:analytic}}
We provide in this Appendix some additional analytical results for the no-jumps evolution. As previously mentioned, this particular case can be thought of as generated by the non-hermitian Hamiltonian in Eq. (\ref{eq:nojump_hamiltonian}), in such a way that a non-normalized state $\ket{\Tilde{\psi}(t)}$ will follow Schrodinger's equation

\begin{equation}
    i\frac{d}{dt}|\Tilde{\psi}(t)\rangle = H_o(t)  \,|\Tilde{\psi}(t)\rangle
    \label{eq:D_sch}
\end{equation}
where $H_o(t)$ is not only non-hermitian but also explicitly time-dependent due to the function $f(t)$. The effective drift Hamiltonian shares eigenstates with $H(t)$, but the eigenvalues associated with these eigenstates are now complex and time-dependent, given by $\pm \omega/2\,\left[1- i\,\Gamma/(2\omega) f(t)\right]$.

The dynamics of the normalized state of the system

\begin{equation}
    \ket{\psi(t)} = \frac{|\Tilde{\psi}(t)\rangle}{\sqrt{\langle\Tilde{\psi}(t)|\Tilde{\psi}(t)\rangle}}
    \label{eq:D_unitary_state}
\end{equation}
will be governed by the more involved, nonlinear equation which is found by jointly differentiating Eq. (\ref{eq:D_unitary_state}) and making use of Eq. (\ref{eq:D_sch}).

The not-normalized state can be expanded into the instantaneous eigenstates of $H_o(t)$, as $\ket{\Tilde{\psi}(t)} = \tilde{c}_{+}\ket{\psi_+(t)} + \tilde{c}_{-}\ket{\psi_-(t)}$. 
Explicit computation of Eq. (\ref{eq:D_sch}) leads to the following differential equations for the coefficients $\Tilde{c}_\pm(t)$
\begin{align}\nonumber
    \Dot{\Tilde{c}}_\pm = \left(\mp i \frac{\omega}{2} - i\frac{\Omega}{2}(1\mp\cos(\theta)) \mp \frac{\Gamma}{4}\,f(t)\right)\Tilde{c}_\pm(t)\\
    +\,i\frac{\Omega}{2}\sin(\theta)\,\Tilde{c}_\mp(t),
    \label{eq:D_diff}
\end{align}
where the real term $\sim -\Gamma \Tilde{c}_+(t)$  indicates that even in the case with no jumps, the presence of the environment favors state transitions, as the amplitude of the excited eigenstate is suppressed. Taking into account the normalization procedure involved in turning from the not-normalized state into the real, normalized one, this suppression implies a population transfer from the excited eigenstate into the ground state. As a consequence, any trivial implementation of the adiabatic approximation is prevented. A second feature observed in Eq. (\ref{eq:D_diff}) is that, for the parameters chosen in this work, a good agreement can be obtained by replacing $f(t)$ with its mean value $f(t)\sim 1 -\sin^2(\theta)/2$. By performing this replacement, dynamics become easily solvable in the rotating frame. The smooth evolution of each eigenstate of the system is, within this approximation, given by
\begin{align}\nonumber
    \ket{\psi_{(\pm)}(t)}& = \mathcal{N}_\pm \,e^{-i\Omega/2\,t}\\ \nonumber
    &\hspace{-1cm}\left\lbrace \left[\pm(\nu+\varepsilon)e^{-i\varepsilon/2\,t}\mp(\nu-\varepsilon)e^{i\varepsilon/2\,t} \right]\ket{\psi_\pm(t)}\right.\\
    &\hspace{2.8cm}\left.-\Omega\sin(\theta)\ket{\psi_\mp(t)}\right\rbrace,
    \label{eq:D_state}
\end{align}
where both $\nu$ and $\varepsilon$ are complex quantities given by $\nu = \omega -\Omega\cos(\theta)- i\,\Gamma/2(1-\sin^2(\theta)/2)$ and $\varepsilon = \sqrt{\nu^2+\Omega^2\sin^2(\theta)}$, 
$\mathcal{N}_\pm$ is a normalization factor. At this point, it should be stressed that Eq.(\ref{eq:D_state}) explicitly shows how the state $\ket{\psi_{(\pm)}(t)}$ obtained when evolving 
an eigenstate will {\em not} be equal to the instantaneous eigenstate at a later time in the general case.  

{\em Geometric phase - } The GP associated with a trajectory in which no jumps can be explicitly computed from Eq. (\ref{eq:GP_nj}). While the general expression is quite involved, it takes, for small rates $\Omega/\omega \sim \Gamma/\omega$ of the driving frequency and the dissipation rate to the amplitude of the magnetic field, the form
\begin{align}
    \phi_0 \sim& -\pi(1-\cos(\theta))\label{eq:D_GP}\\[.75em]\nonumber  &- \pi\sin^2(\theta)\left(\frac{\Omega}{\omega} + \cos(\theta)\frac{\Omega^2}{\omega^2}\right)\\[.75em]\nonumber
    &- \frac{\sin^2(\theta)}{4}\left(\frac{\Omega}{\omega} + \cos(\theta)\frac{\Omega^2}{\omega^2}\right)\frac{e^{-4\pi\Im(\nu)/\Omega}-1}{2\Im(\nu)/\Omega},
\end{align}
where the first term in the r.h.s is the Berry phase. The term in the second line of the equation is the main correction originating exclusively from non-adiabaticity, in otherwise unitary evolution. The third line accounts for the non-trivial effect of the environment in the no-jump evolution. As $\Gamma\rightarrow 0$ this term turns into a further contribution due to non-adiabaticity.  

{\em Phase diagram singularities - } When computing the accumulated GPs analyzed in Sections \ref{sec:unrav_gp} and \ref{sec:topological} we have taken $\ket{\psi_+(0)}$ as our 
initial state. Thus, a vanishing probability for observing this particular trajectory, of the kind observed at the GP singular points, requires $\ket{\psi(T)} \sim \ket{\psi_-(T)}$. 
Considering the cyclic character of the instantaneous eigenstates, this means a singular point will take place whenever a full population transfer is achieved exactly in a time period.
It was already inferred from the differential equations governing the evolution of the $\Tilde{c}_\pm$ coefficients, that the dynamics generated by the effective drift Hamiltonian $H_o(t)$ favored transitions from the excited to the ground instantaneous eigenstate. As long as the original approximation remains accurate, the singular points of the GP will be defined through the equation $(\nu+\varepsilon)-(\nu-\varepsilon) e^{2i\pi\varepsilon/\Omega} = 0$.

{\em No-jump interference fringe - } In Section \ref{sec:unrav}, we have studied the interference fringes of an echo experiment. For this purpose, we've defined the convenient parameter $\varphi$ given by Eq. (\ref{eq:xdefinition}). Restricting to the case of a protocol performed without registering any jump, it was shown that the value of $\varphi$ displays, generally, better agreement with the Berry phase than with the GP $\phi_0$ accumulated by the state of the system under equal conditions, this is, when it is smoothly driven along one period of time. As long as the no-jump value $\varphi \sim \phi_{\rm a}$, good agreement between this “phase" and the GP will be obtained when the second and third lines in Eq. (\ref{eq:D_GP}) are sufficiently small. However, it is worth noting that the $\varphi$ value will not remain close to the Berry phase for arbitrarily small driving frequency. While the  protocol has shown to be less sensitive to both non-adiabatic and environmentally induced effects than the GP, it will account for the non-ideal conditions. It was already shown that the environment induces population transfer from the excited to the ground state. The asymmetry between the smooth evolution of each eigenstate should be expected to prevent, at some point, the cancellation of the dynamical evolutions.  Figure \ref{fig:varphi_deviation} illustrates this situation, by showing the $\varphi$ value as a function of the rate between the driving frequency and the field amplitude. While for larger rates $\gtrsim 0.01$ the phase reproduces the behavior discussed in Section \ref{sec:unrav}, this situation does not hold if the rate is lowered enough. 
At some critical value, the parameter extracted from the echo protocol starts deviating from the adiabatic phase.  

A rather singular situation arises when the state at the end of the protocol coincides, up to a global phase, with $\ket{\psi_-(0)}$, so that the persistence probability turns $\mathcal{P}=1/2$. 
As a consequence, the $\varphi \sim 1.375\pi$ value observed in Fig. \ref{fig:varphi_deviation}, trivially associated with $\mathcal{P}\sim 1/2$ by Eq. (\ref{eq:xdefinition}) is obtained. In this case, the three peaks observed in the distribution $\mathcal{P}[\varphi]$ (see Section \ref{sec:unrav}) merge into a single, central peak. This regime is therefore accessed when full population transfer occurs within a cycle and the system reaches a steady state $\sim \ket{\psi_-(t)}$. 
As we have discussed above, the parameters leading to full population transfer exactly in a cycle $t\in[0,T]$ correspond to singular points of the GP. Then, full population transfer {\em within} a cycle implies evolutions performed either at higher dissipation rates or at slower frequencies than those for which the singularity occurs. This requirement establishes a connection between the value of the echo phase and the topological classes of evolution, as distinctive regimes of $\varphi$ are accessed on one and the other side of the singular points. 

\bibliographystyle{unsrtnat}
\bibliography{cites.bib}
\end{document}